\documentclass{jfm}

\usepackage{epsfig}
\usepackage{natbib}
\usepackage{color}
\usepackage{amsmath}
\usepackage{amssymb}
\usepackage{upmath}

\ifCUPmtlplainloaded \else
  \checkfont{eurm10}
  \iffontfound
    \IfFileExists{upmath.sty}
      {\typeout{^^JFound AMS Euler Roman fonts on the system,
                   using the 'upmath' package.^^J}%
       \usepackage{upmath}}
      {\typeout{^^JFound AMS Euler Roman fonts on the system, but you
                   dont seem to have the}%
       \typeout{'upmath' package installed. JFM.cls can take advantage
                 of these fonts,^^Jif you use 'upmath' package.^^J}%
      }
  \else
  \fi
\fi

\ifCUPmtlplainloaded \else
  \checkfont{msam10}
  \iffontfound
    \IfFileExists{amssymb.sty}
      {\typeout{^^JFound AMS Symbol fonts on the system, using the
                'amssymb' package.^^J}%
       \usepackage{amssymb}%

      }{}
  \fi
\fi

\ifCUPmtlplainloaded \else
  \IfFileExists{amsbsy.sty}
    {\typeout{^^JFound the 'amsbsy' package on the system, using it.^^J}%
     \usepackage{amsbsy}}
    {\providecommand\boldsymbol[1]{\mbox{\boldmath $##1$}}}
\fi

\newcommand\Rey{\mbox{\textit{Re}}}  

\newcommand{\Weber}{\mbox{\textit{We}}}  
\newcommand{\Bond}{\mbox{\textit{Bo}}}   
\newcommand{\Froude}{\mbox{\textit{Fr}}} 

\newsavebox{\astrutbox}
\sbox{\astrutbox}{\rule[-5pt]{0pt}{20pt}}

\newcommand\squart{\ensuremath{{\textstyle\frac{1}{4}}}}
\newcommand\thalf{\ensuremath{{\textstyle\frac{1}{2}}}}

\graphicspath{{figures/}}

\title[Splash wave and crown breakup after disc impact on a liquid surface.]{Splash wave and crown breakup after disc impact on a liquid surface.}

\author[Ivo R. Peters, Devaraj van der Meer and Jos\'e Manuel Gordillo]
{I\ls V\ls O\ns R.\ns P\ls E\ls T\ls E\ls R\ls S\ls$^1$, \ns
D\ls E\ls V\ls A\ls R\ls A\ls J\ns v\ls a\ls n\ns d\ls e\ls r\ns M\ls E\ls E\ls R\ls$^1$
\and
J.\ns M.\ns G\ls O\ls R\ls D\ls I\ls L\ls L\ls O\ls$^{2}$\thanks{Corresponding author: jgordill@us.es}\break}
\affiliation{$^1$ Department of Applied Physics and J.M. Burgers
Centre for Fluid Dynamics, University of Twente, P.O. Box 217,
7500 AE Enschede, The Netherlands\\
$^2$ \'Area de Mec\'anica de Fluidos, Departamento de Ingener\'ia
Aeroespacial y Mec\'anica de Fluidos, Universidad de Sevilla,
Avenida de los Descubrimientos s/n 41092, Sevilla, Spain.}

\pubyear{????}
\volume{???}
\pagerange{???--???}
\date{?? and in revised form ??}

\begin{document}

\maketitle

\begin{abstract}
In this paper we analyze the impact of a circular disc on a free surface using experiments, potential flow numerical simulations and theory. We focus our attention both on the study of the generation and possible breakup of the splash wave created after the impact and on the calculation of the force on the disc. We have experimentally found that drops are only ejected from the rim located at the top part of the splash --giving rise to what is known as the crown splash-- if the impact Weber number exceeds a threshold value $\Weber_{crit}\simeq 140$.
We explain this threshold by defining a local Bond number $Bo_{tip}$ based on the rim deceleration and its radius of curvature, with which we show using both numerical simulations and experiments that a crown splash only occurs when $Bo_{tip}\gtrsim 1$, revealing that the rim disrupts due to a Rayleigh-Taylor instability.
Neglecting the effect of air, we show that the flow in the region close to the disc edge possesses a Weber-number-dependent self-similar structure for every Weber number. From this we demonstrate that $\Bond_{tip}\propto\Weber$, explaining both why the transition to crown splash can be characterized in terms of the impact Weber number and why this transition occurs for $We_{crit}\simeq 140$.

Next, including the effect of air, we have developed a theory which predicts the time-varying thickness of the very thin air cushion that is entrapped between the impacting solid and the liquid. Our analysis reveals that gas critically affect the velocity of propagation of the splash wave as well as the time-varying force on the disc, $F_D$. The existence of the air layer also limits the range of times in which the self-similar solution is valid and, accordingly, the maximum deceleration experienced by the liquid rim, what sets the length scale of the splash drops ejected when $We>\Weber_{crit}$.
\end{abstract}


\section{Introduction}
\label{sec:introduction}

The seemingly straightforward experiment of impacting a solid or liquid object on a liquid surface exhibits many challenges for physical understanding, as was already noticed more than a century ago \citep{Worthington1896,Worthington1900,Worthington1908}. The creation and collapse of an underwater cavity has been studied extensively for a solid impacting a liquid surface \citep{Richardson1948,Gaudet1998,Lee1997,Bergmann2006,Ducleaux2007,Bergmann2009}, as well as the formation of jets resulting from the collapse of this cavity \citep{Longuet-higgins1983,Hogrefe1998,Zeff2000,Duchemin2002,Gekle2009a,Gekle2010d,Gordillo2010}. The first event that is visible after an object hits a liquid is however the splash. It is formed by the liquid that is moving upwards close to the downwards moving object, which initially can be considered as a Wagner problem, under the condition that the object can locally be approximated as flat \citep{Wagner1932,Scolan2001}. When the object is not locally flat, a splash will only be formed provided that certain conditions are fulfilled. These conditions can sometimes be as subtle as the wetting properties of a smooth sphere, as was shown by \cite{May1951,Duez2007}.

The impact of a flat plate on a free surface allows for analytical investigations and can also be studied using experiments and numerical simulations. Self-similar solutions for the case without surface tension ($\Weber\rightarrow\infty$) have been found by \cite{Iafrati2004} for the initial stage after impact, whose scaling exponents were already noticed by \cite{Yakimov1973}.
Later, this analysis was expanded in order to calculate the hydrodynamic load on a flat surface close to impact by \cite{Iafrati2008,Iafrati-Korobkin11}. One of the existing differences between our study and the above mentioned ones is that we retain surface tension effects in the analysis. We show that self-similar solutions exists for \emph{any} value of the Weber number, a fact corroborated with the aid of boundary integral (BI) simulations. We extend our analysis by accounting for air effects, and show that these enormously influence both the splash wave velocity during the initial instants after impact and the force on the disc. Surprisingly, the gas-to-liquid density ratio also determine the sizes of the drops generated when the impact speed is above a certain threshold. Indeed, the liquid that is thrown upwards due to the impact of an object, develops in a thin sheet with a cylindrical rim on top of it \citep{Yarin2006} which is susceptible to instabilities that may result in the ejection of droplets. Finding the nature of instabilities that result in the formation of droplets has been the motivation for many studies, examples of which are given below.

In \cite{Krechetnikov2009a}, it is argued that the crown formation is the result of a Richtmyer-Meshkov instability, and the effect of interfacial curvature on liquid rims is discussed in \cite{Krechetnikov2009}, again in the light of Richtmyer-Meshkov and Rayleigh-Taylor instabilities.
\cite{Deegan2008} showed that in a narrow range of parameters, this instability can develop in a very regular pattern, which was later used by \cite{Zhang2010a} to experimentally show that the wavelength of this regular pattern agrees with the predicted value corresponding to a Rayleigh-Plateau instability. In an analytical study, \cite{Krechetnikov2010} elaborates more on the interplay between the Rayleigh-Plateau and the Rayleigh-Taylor instability. Recently, \cite{Lister2011} specifically isolated the Rayleigh-Taylor instability on a cylinder, tuning the body force normal to its surface by tilting a liquid cylinder inside a liquid of higher density. Finally, \cite{Lhuissier2011} found that a Rayleigh-Taylor instability is responsible for the formation of ligaments from the liquid rim observed in bursting bubbles.

In this paper we argue that the splash wave deceleration on its top part plays a key role in the ejection of droplets. Indeed, we find that drops will be detached when the 
effective weight per unit length of the liquid in the rim, $\rho(A_{tip}-g)\,R^2_c$, overweighs the surface tension forces per unit length $\sigma$, namely, when the local Bond number satisfies the condition
\begin{equation}
    \Bond_{tip}=\rho(A_{tip}-g)R_C^2/\sigma>1\, ,
    \label{eq:localBond}
\end{equation}
where $A_{tip}$ is the rim deceleration, $\rho$ and $\sigma$ are, respectively, the density and surface tension of water and $R_C$ is the radius of curvature of the rim. We combine experiments, numerical simulations and theory to show that the local Bond number at the rim is the relevant parameter to predict the transition to crown splash.

The paper is organized as follows. In \S\ref{sec:experiments} we experimentally describe the different spatial regions appearing after the impact of a solid disc against a free surface and show that the experimental results can be reproduced by using potential flow numerical simulations. Making use of potential flow theory, we find in \S\ref{sec:SplashTheory} that, when air effects are absent, the splash region possesses a self similar structure that depends only on \Weber\ for dimensionless times satisfying $t\ll 1$. In section \S\ref{sec:splashTransition} we find the threshold value for the critical Weber number $\Weber_{crit}$ above which the crown breaks up and droplets are ejected. In \S\ref{sec:air} we reveal the critical role played by air density in setting the maximum deceleration of the rim and, using this information, we deduce an expression for the drop size when \Weber$>\Weber_{crit}$. In addition, in this section we study the force exerted on the disc during the impact process and show how it is regularized by the air layer in between the disk and the liquid. Conclusions are drawn in \S\ref{sec:conclusions}.


\section[Experiments and comparison with BI simulations]{Experiments and comparison with boundary integral simulations.}
\label{sec:experiments}

Our experimental setup consists of a disc with radius $R_D$ which we pull down through a water surface at a constant speed $V_D$ using a linear motor. The linear motor ensures that our disc always moves with a constant prescribed velocity. We record the events using a Photron SA1 high speed camera. A more detailed description of the experimental setup can be found in \cite{Bergmann2009}. From now on, distances, times, velocities and pressures are made dimensionless using $R_D$, $R_D/V_D$, $V_D$ and $\rho V^2_D$ as characteristic length, time, velocity and pressure respectively. Also, variables in capital letters will denote dimensional quantities, whereas their corresponding dimensionless counterparts will be expressed using lower case letters.

\begin{figure}
    \centering
    \includegraphics[width=12cm]{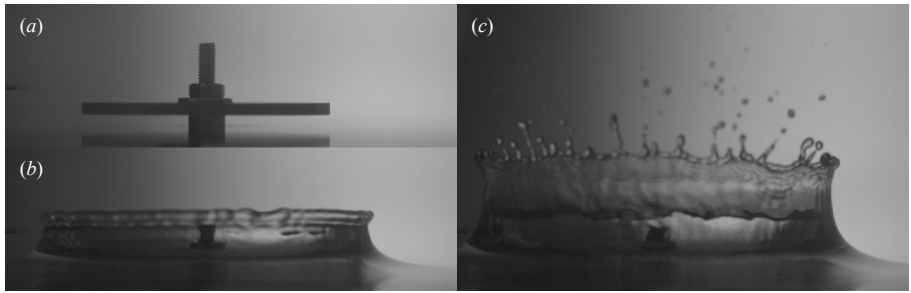}
    \caption{The splash generated at $T=0.01~\mathrm{s}$ after the normal impact against a water interface of a circular disc with $R_D=20~\mathrm{mm}$. ($a$) A snapshot of the disk just before the impact takes place. ($b$) The splash created by an impact below the critical Weber number impact velocity $V_D=0.6~\mathrm{m/s}$; $\Weber=99$). No drops are ejected from the tip of the rim since $\Weber<\Weber_{crit}$. ($c$) The splash created by an impact above the critical Weber number
($V_D=1.0~\mathrm{m/s}$; $\Weber=274$). Since $\Weber>\Weber_{crit}$, the rim breaks into drops with sizes of the order of the rim width.
}
    \label{fig:Exp0}
\end{figure}
Figure \ref{fig:Exp0} shows the surface deformation that occurs immediately after a circular disc impacts perpendicularly with a constant velocity onto a free surface bounding a deep water layer. From these images, it can be appreciated that a circular liquid sheet --the splash-- is ejected out of the liquid bulk all along the perimeter of the disc. The sheet then propagates radially outwards, `informing' the rest of the fluid of the solid body impact. The liquid speeds inside the splash are much larger than the disc impact velocity $V_D$: indeed, for a given instant in time, the distance traveled out of the liquid bulk by the top part of the liquid sheet is much larger than the distance traveled within the liquid by the disc. We observe that, depending on the impact velocity, the liquid sheet can either breakup into drops or just retract into the liquid bulk without breaking. Indeed, if $V_D$ or, equivalently, if the impact Weber number $\Weber=\rho R_D V_D^2/\sigma$ is sufficiently large, the rim at the highest part of the sheet breaks into drops, provoking what we refer to as the crown breakup or crown splash [figure~\ref{fig:Exp0}($c$)]. If, in contrast, $V_D$ (or \Weber) is sufficiently small, surface tension forces and gravity pull back the edge of the rim into the liquid and no breakup occurs [figure~\ref{fig:Exp0}($b$)].

\begin{figure}
    \centering
    \includegraphics[width=12cm]{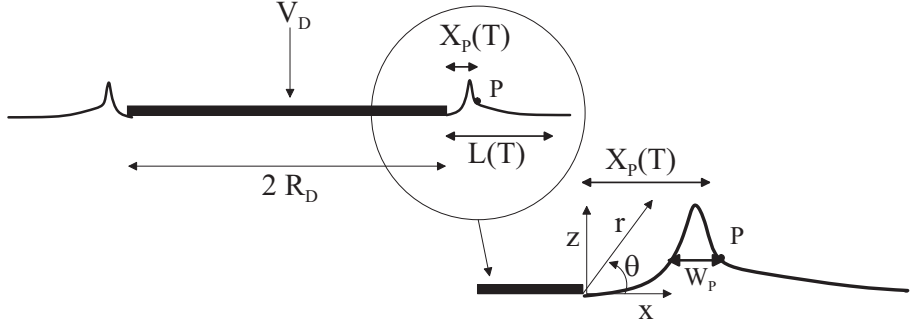}
    \caption{Schematic drawing of the impact of the disc and generation of the splash wave. We define the origin of our coordinate system at the edge of the disc for both the cartesian and polar coordinate system. The point P, which we use as a reference point to measure velocities and lengths in the splash, is defined where the slope of the splash wave is $-1$. The length of the splash region, $X_P(T)$, is the distance between the disc edge and the point P. This variable is used to indicate at which distance from the edge of the disc the free surface has deformed significantly. We define the intermediate length $L(T)$, such that $L(T)\gg X_P(T)$, but small enough when compared to $R_D$ to approximate the flow near the disc edge as 2-dimensional, as explained in \S\,\ref{sec:SplashTheory}.}
    \label{fig:Esquema}
\end{figure}

\begin{figure}
    \centering
    \includegraphics[width=\textwidth]{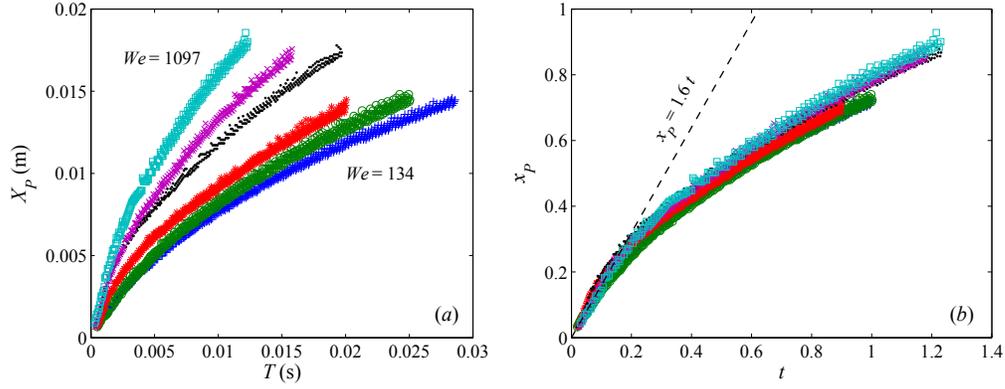}
    \caption{($a$) Position of the point P defined in figure \ref{fig:Esquema} as a function of time for several values of the disc impact velocity. ($b$) The same curves as in ($a$) but expressed in dimensionless units. Observe that, initially, the speed of the splash wave is constant and proportional to $V_D$ since, initially, $\dot{x}_P\simeq 1.6$. After a short time interval, the wave speed is no longer constant but decreases in time.}
    \label{fig:Xs}
\end{figure}

Note in figures~\ref{fig:Exp0}--\ref{fig:Esquema} that initially the splash wave develops in a confined spatial region, which grows in time.  More specifically, this region has a characteristic length $X_P(T)\ll R_D$ located very close to the disc edge, with $X_P$ the distance from the disc edge to the point $P$ indicated in figure \ref{fig:Esquema} where the slope in the splash is -1. Figure \ref{fig:Xs}($a$) shows the distance traveled by the point P in the splash wave as a function of time for several values of the impact velocity. Clearly, the propagation velocity of the splash wave increases as the impact velocity of the disc increases. The different curves in figure \ref{fig:Xs}($a$), when represented using dimensionless units, collapse onto a single one as it is depicted in figure \ref{fig:Xs}($b$), meaning that the dimensionless velocity of the splash wave is independent of the impact Weber number. The results in figure \ref{fig:Xs}($b$) also indicate that the speed of the splash wave is initially constant since $\dot{x}_P\simeq 1.6\rightarrow \dot{X}_P\simeq 1.6\,V_D$ for $t\simeq 0$. However, after a short time interval, the splash velocity is no longer constant but decreases in time.

Due to the fact that the impact Reynolds number $\Rey=V_D R_D/\nu$ (where $\nu$ indicates the kinematic viscosity of the liquid) is such that $\Rey\gtrsim O(10^4)$, we expect that viscous effects are confined to a thin boundary layer developing at the disc bottom. Thus, the velocity field in most of the liquid volume is expected to be described using a velocity potential. As a consequence of this, the time evolution of the free surface will be correctly predicted using a potential flow boundary integral method of the type used to describe the collapse of cavities \citep{Longuet-Higgins1995,Gekle2008}, the ejection of Worthington jets \citep{Gekle2009a,Gekle2010d,Gordillo2010} or the formation of bubbles from an underwater nozzle \citep{Oguz1993}.

\begin{figure}
    \centering
    \includegraphics[width=12cm]{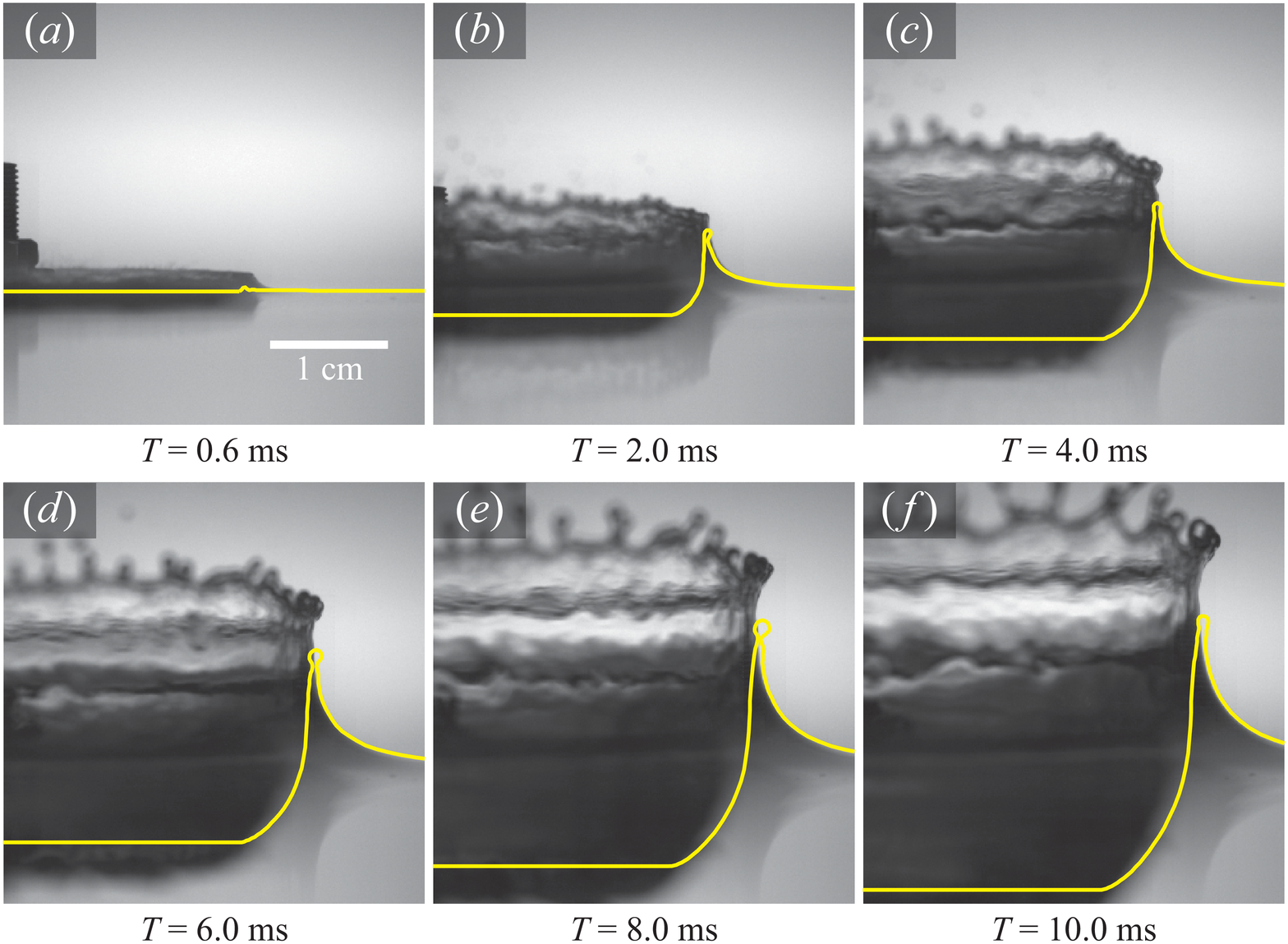}
    \caption{Six snapshots of the experiment at different times $T$ after disc impact, compared with the corresponding results from the boundary integral simulations (yellow lines). Disc radius is $20~\mathrm{mm}$, the impact speed is $1~\mathrm{m/s}$. The agreement between numerical simulations and experiments is excellent once two precisions are made: First, in the numerical simulations, the tip of the splash is unstable and therefore breaks earlier than in experiments. For this reason, the splash in the experiment appears higher than in the simulations. Second, the times corresponding to the experimental profiles are those of the numerical ones plus a constant $T_{0exp}\simeq0.6~\mathrm{ms}$. The existence of this time shift is related to the influence of air at the bottom of the disc and can be retained by simply shifting the numerical origin of times by a constant value (virtual origin of the self similar solution, see text).}
    \label{fig:Exp}
\end{figure}

Indeed, figure~\ref{fig:Exp} shows a single fluid potential flow simulation of a disc impacting at a constant velocity $V_D$ against a free surface which very well reproduces the experiment once the origin of times is shifted in time by a quantity $t_{0exp}$. It will be shown below that $t_{0exp}$ plays the role of a virtual origin in time which accounts for the presence of air between the bottom of the plate and the free surface, as will be discussed in \S\ref{sec:air}.  We conclude from the result depicted in figure \ref{fig:Exp} that the liquid flow can be accurately reproduced for $t\gtrsim t_{0exp}$ using a single fluid potential flow description.

\section{Theoretical description of the splash}
\label{sec:SplashTheory}

In this section we will provide a theoretical description for the formation of the splash neglecting the effect of air density, which is in part analogous to \cite{Iafrati2004}. Let us first notice from figures ~\ref{fig:Exp0}--\ref{fig:Esquema} that there exist two well-differentiated spatial regions after a solid impacts a free surface. Indeed, using the (dimensional) polar coordinates $(R,\theta)$ centered at the disc edge shown in figure~\ref{fig:Esquema}, we observe that there exists an inner region $\sim X_P(T)\ll R_D$, where the interface deforms appreciably and an outer region $X_P(T)\ll L(T)\ll R_D$ where the interface hardly changes from its initial position. We start with deriving the flow field in the outer region, i.e., at distances $r$ from the edge of the disc such that $x_P\ll r\ll 1$, for times close to the moment of impact ($t\ll1$), neglecting air effects and assuming negligible deformations of the free surface (Section \ref{subsec:flowfield}). After that, we use the analytical expression of this flow field as the far field boundary condition for the inner problem: The flow within the splash region, $r\sim x_P$, where the deformations of the free interface are no longer small. We will show theoretically that this specific type of far field boundary condition leads to the flow field within the splash region to be self-similar. We confirm this result by rescaling the splash wave profiles computed using the numerical boundary integral method for several values of the impact Weber number (Section \ref{subsec:selfsim}). Finally, we show that the expected self-similar scalings related to the vertical positions and vertical velocities are fully recovered after we compensate for the downwards motion of the disc, which introduces a non self-similar correction (Section \ref{subsec:selfsimcorrection}).

\subsection{Flow field}
\label{subsec:flowfield}
Since $x_P(t)\ll 1$, it is possible to define an intermediate length $\ell(t)$ such that $x_P(t)\ll \ell(t)\ll 1$ at which the height of the interface hardly varies with respect to its initial position (see figure \ref{fig:Esquema}). Since $\ell(t)\ll 1$, the velocity potential $\phi$ at the intermediate region can be described using a two dimensional approach, which satisfies the following equation:
\begin{equation}
    \nabla^2\phi = \frac{\partial^2\phi}{\partial x^2}+\frac{\partial^2\phi}{\partial z^2}=0\, , \label{Laplace2D}
\end{equation}
subject to the following boundary conditions:
\begin{equation}
    \frac{\partial \phi}{\partial z}=-1\quad \mathrm{at}\quad z=-t\simeq 0\,\,; \,\,\, x<0\, ,\label{eq:BCDisc}
\end{equation}
which is the kinematic boundary condition imposed by the downward moving disc,
\begin{equation}
    \phi\simeq0  \quad \mathrm{at}\quad z\simeq 0\,\,; \,\,\,  x>0 \, ,
    \label{eq:BCFreeSurface0}
\end{equation}
denoting the dynamic boundary condition at the free surface\footnote{Equation (\ref{eq:BCFreeSurface0}) has been obtained by approximating the time-integrated unsteady Bernoulli equation for very small values of $t$, which for $z=0$ and $x>0$ can be written as $\phi\simeq\phi(t=0)-\thalf\left|\nabla\phi(t=0)\right|^2t\simeq \phi(t=0)$, and taking into account that $\phi(t=0)=0$ for $z=0$ and $x>0$.}, and
\begin{equation}
    \phi\rightarrow 0 \quad \mathrm{for}\quad \sqrt{x^2+z^2}\rightarrow \infty\, ,
    \label{BCInfty0}
\end{equation}
enforcing the fluid far away from the impact to be at rest.

\begin{figure}
    \centering
    \includegraphics[width=10cm]{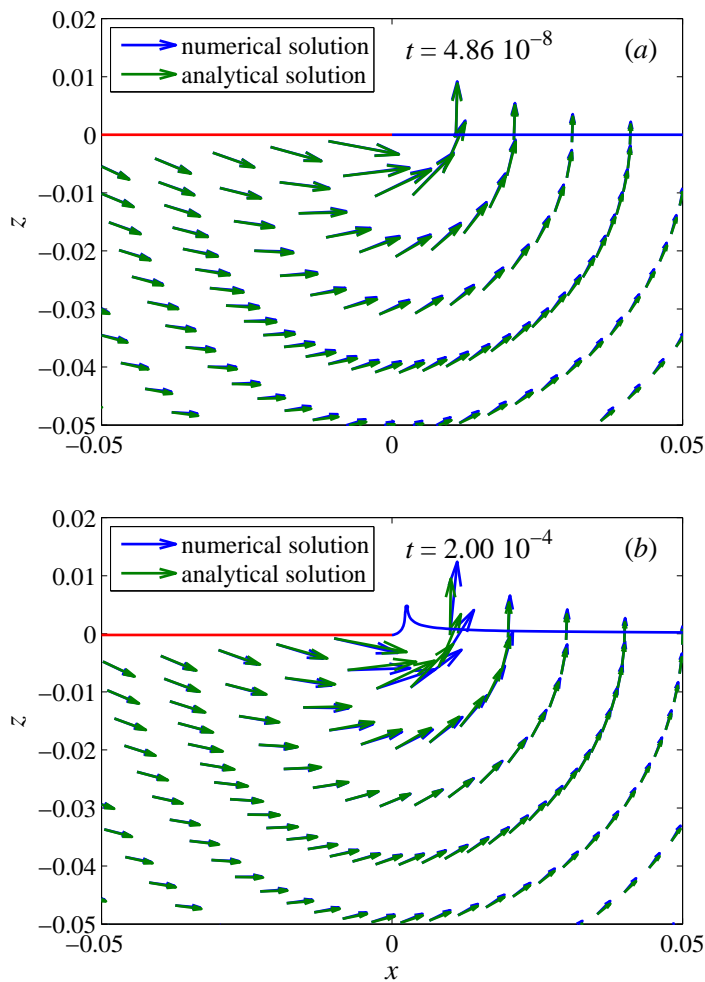}
    \caption{The analytical flow field given in equation~(\ref{eq:VelocityField}) compares very favorably with the numerical one once the constant $A$ is set to $0.44$. (\emph{a}) At an extremely short time after impact, when the free surface has not yet deformed, the analytical solution agrees with the boundary integral result in the full inner domain where $r\ll1$. (\emph{b}) At a later point in time we observe that close to the splash region, where the deformation is appreciable, the analytical flow field deviates from the numerical solution. Away from the splash region, the agreement improves again.
    }
    \label{fig:flowField}
\end{figure}

Here, we used the cartesian coordinates $(x,z)$ defined in figure~\ref{fig:Esquema} and have taken into account the observations in figure~\ref{fig:Exp}: that for $t\ll1$ ($T\ll20~\mathrm{ms}$ in figure~\ref{fig:Exp}), the splash develops close to the edge of the disc and that the interface is not appreciably distorted in the intermediate region $r\sim \ell(t)$. The solution to the system (\ref{Laplace2D})-(\ref{BCInfty0}) can be found using standard 
conformal mapping techniques, yielding the complex flow field
\begin{equation}
    \frac{d\omega}{d\zeta} \equiv \frac{\partial\phi}{\partial x}-i\,\frac{\partial\phi}{\partial z}=i+i\,A\,r^{-1/2}e^{-i\theta/2}\left(1+\thalf re^{i\theta}\right)\left(1+\squart r e^{i\theta}\right)^{-1/2} \, ,
    \label{eq:VelocityField}
\end{equation}
where $i=\sqrt{-1}$ and the constant $A$ is determined by matching the velocity field given in equation~(\ref{eq:VelocityField}) with the numerical solution of the tangential velocity field at the disc edge, which takes into account the true, i.e., three-dimensional, geometry of the impactor and the corresponding flow. Figure~\ref{fig:flowField} shows a comparison for two instants of time, such that $t\ll 1$, between the numerical calculated velocity field and the one given by equation~(\ref{eq:VelocityField}), which is valid in the intermediate region $r\sim\ell(t)$. The agreement is almost perfect when the deformation of the free surface is minimal, as in figure~\ref{fig:flowField}(\emph{a}). Figure~\ref{fig:flowField}(\emph{b}) shows the influence of the splash region $r\sim x_P$, where there is a clear discrepancy between the approach of equation~(\ref{eq:VelocityField}) and the numerical solution. However, looking at the region $r\sim\ell(t)$, where deformation can be neglected, the agreement is fully recovered.

Now, since the solution (\ref{eq:VelocityField}) constitutes the outer boundary condition for the velocity field in the inner region $r\sim x_P(t)\ll\ell(t)\ll1$, we are interested in the approximation of this equation in the limit $r\ll1$, which yields,
\begin{equation}
    \frac{\partial\,\phi}{\partial\,x} \simeq
    A r^{-1/2}\sin(\theta/2)\, ,\quad
    \frac{\partial\,\phi}{\partial\,z} \simeq     -1-Ar^{-1/2}\cos(\theta/2) \label{Veledge}
\end{equation}
and the potential
\begin{equation}
    \phi \simeq -z-2Ar^{1/2}\sin(\theta/2)\, , \label{potencialdelinfinito}
\end{equation}
which will serve as a boundary condition to the problem of determining the free surface shape and potential in the splash region $r\sim x_P(t)$. Notice from the first equation in (\ref{Veledge}) that the modulus of the velocity component tangent to the disc bottom surface ($\theta=\pi$) diverge as $r^{-1/2}$. We will show that the huge tangential velocities near the disc edge, which have to accommodate to the mass of fluid at rest, are responsible for the generation of the splash wave.

\subsection{Self similarity}
\label{subsec:selfsim}
In the frame of reference moving at the disc velocity, the splash region $r\sim x_P(t)\ll\ell(t)$ can be described by solving the Laplace equation (\ref{Laplace2D}) subjected to  the following boundary conditions at a given instant in time:
\begin{equation}
    \frac{\partial \phi}{\partial z}=0\quad \mathrm{at}\quad z=0 \quad x<0\, ,
    \label{eq:BCDiscFrame}
\end{equation}
which is the kinematic boundary condition at the bottom of the disc, and
\begin{equation}
    \frac{\partial\phi}{\partial t}+\frac{|\nabla\phi|^2}{2}+\frac{\kappa}{\Weber}+\frac{z}{\Froude}=0 \quad \mathrm{at}\quad z=f(x,t)\quad x>0\, ,
    \label{Bernoulli}
\end{equation}
which is the dynamic boundary condition at the free surface, with $\Weber=\rho V^2_D R_D/\sigma$ the Weber number and $\Froude=V_D^2/(gR_D)$ the Froude number. These equations need to be complemented with the far field velocity potential in the region where the interface is virtually undisturbed, i.e.,
\begin{equation}
    \phi\rightarrow -2Ar^{1/2}\sin(\theta/2)\quad \mathrm{for}\quad r\rightarrow \infty\, .
    \label{eq:BCInfty}
\end{equation}
The function $f(x,t)$ defines the position of the free interface, which satisfies the kinematic boundary condition
\begin{equation}
    \frac{\partial\,f}{\partial \,t}=\frac{\partial\phi}{\partial z}-\frac{\partial\phi}{\partial x}\frac{\partial f}{\partial x}\quad \mathrm{at}\quad z=f(x,t) \quad x>0 \, ,
    \label{Free}
\end{equation}
where
\begin{equation}
    f(x,t=0)=0 \quad \mathrm{and}\quad f(x\rightarrow \infty,t)\rightarrow t.
    \label{eq:FreeInfty}
\end{equation}
Note that in the Bernoulli equation (\ref{Bernoulli}) we have used the interfacial curvature
\begin{equation}
    \kappa=\frac{\partial^2 f}{\partial x^2}
    \left( 1 + \left( \frac{\partial f}{\partial x} \right)^2 \right)^{-3/2}
\end{equation}
to express the pressure jump across the surface, i.e., $p=\kappa\Weber^{-1}$. Since there is no characteristic length scale in the system of equations (\ref{Laplace2D}) and (\ref{eq:BCDiscFrame})-(\ref{Free}), we expect the existence of self-similar solutions of the type
\begin{equation}
    \phi = (t-t_0)^{\beta}\bar{\phi}\left(\frac{x}{(t-t_0)^{\alpha}},
    \frac{z}{(t-t_0)^{\alpha}}\right),
    \label{eq:Self1}
\end{equation}
which we now write as
\begin{equation}
    \phi = \tau^{\beta}\bar{\phi}(\chi,\eta),
    \label{eq:Self2}
\end{equation}
with $\chi=x/\tau{^\alpha}$, $\eta=z/\tau{^\alpha}$, and $\tau=t-t_0$. The dimensionless quantity $t_0$ can be an arbitrary constant, and the shape of the free surface can be expressed as
\begin{equation}
    f(x,t)= \tau^{\alpha}F(\chi)\, .
    \label{eq:Self3}
\end{equation}

From the Bernoulli equation (\ref{Bernoulli}) we find, by comparing the first two terms, that self-similar solutions can only exist if $\beta=2\alpha-1$. Moreover, the matching with the boundary condition at infinity (\ref{eq:BCInfty}) imposes the additional restriction $2\beta=\alpha$. Combining these two conditions then results in $\beta=1/3$ and $\alpha=2/3$. Thus, lengths are expected to scale with $\tau^{2/3}$ and velocities with $\tau^{-1/3}$. Indeed, the system of equations that we solve for both $\bar{\phi}$ and $F$ reads, with relative errors $\sim O(\tau^{1/3})\ll 1$,
\begin{equation}
    \frac{\partial^2\bar{\phi}}{\partial \chi^2}+\frac{\partial^2\bar{\phi}}{\partial \eta^2}=0\, ,
    \label{Laplace2DSS}
\end{equation}
\begin{equation}
    \frac{\partial\bar{\phi}}{\partial\eta}=0\quad \mathrm{at}\quad \eta=0\, ,\chi<0\,
\end{equation}
\begin{equation}
    \bar{\phi}\rightarrow -2A\bar{r}^{1/2}\sin(\theta/2)\quad \mathrm{for}\quad \bar{r}\rightarrow\infty\, ,\quad \mathrm{with}\quad \bar{r}=\sqrt{\eta^2+\chi^2}\label{BCInftySS}
\end{equation}
\begin{equation}
    \begin{split}
    & \frac{1}{3} \bar{\phi}-\frac{2}{3} \left(\chi\frac{\partial\bar{\phi}}{\partial\chi}+\eta\frac{\partial\bar{\phi}}{\partial\eta}\right) + \thalf\left[\left(\frac{\partial\bar{\phi}}{\partial\chi}\right)^2 + \left(\frac{\partial\bar{\phi}}{\partial\eta}\right)^2\right] + \frac{\bar{\kappa}}{\Weber} + \frac{\eta \tau^{4/3}}{\Froude}=0\\ & \mathrm{at}\quad \eta=F(\chi) \, ,
    \end{split}
    \label{eq:BernoulliSS}
\end{equation}
with
\begin{equation*}
    \bar{\kappa} = \frac{d^2F}{d\chi^2}\left(
    1 + \left( \frac{dF}{d\chi} \right)^2
    \right)^{-3/2}\, ,
\end{equation*}
and
\begin{equation}
    \frac{2}{3} F-\chi\frac{d F}{d\chi}=\frac{\partial\bar{\phi}}{\partial\eta}-\frac{d F}{d\chi}\frac{\partial\bar{\phi}}{\partial\chi}\, \quad\mathrm{at}\quad \eta=F(\chi)\quad \mathrm{and}\quad F(\chi)\rightarrow \frac{\tau+t_0}{\tau^{2/3}}\quad\mathrm{for}\quad \chi\rightarrow \infty \, .
    \label{FreeSS}
\end{equation}
Note that there are two terms breaking the self-similarity of equations (\ref{Laplace2DSS})-(\ref{FreeSS}) namely, the last term of equation (\ref{eq:BernoulliSS}) and the far field boundary condition in equation (\ref{FreeSS}) due to the presence of $\tau$ in them. There will therefore exist a self-similar solution whenever $\tau\gtrsim t_0$ and both $\tau^{1/3}\ll 1$ and $\tau^{1/3}\ll \Froude^{1/4}$ ($\sim O(1)$) are satisfied. In other words, assuming that $t_0$ is sufficiently small, the self similar solution will be valid during a short time interval after impact. Under the conditions in which the flow field is self similar, the disc has moved down only slightly, the normal velocity boundary condition is then approximately applied at the undisturbed free surface and gravitational effects can be neglected.

\begin{figure}
    \centering
    \includegraphics[width=\textwidth]{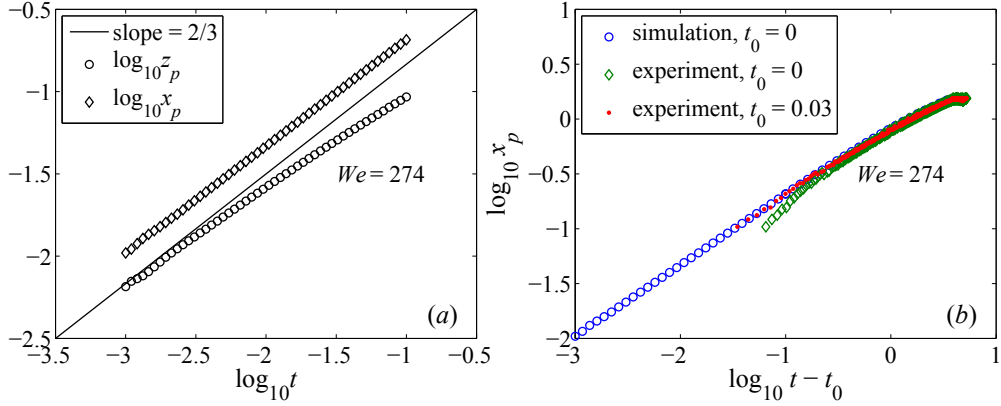}
    \caption{(\emph{a}) Time evolution of the horizontal and vertical coordinates of point P. Numerical results reveal that the scaling for both $x_p$ and $z_p$ follows the prediction of equation equation~(\ref{eq:Self1}) for $\alpha=2/3$. As explained in the main text, the results for $z_p$ slightly deviate from a pure power law due to the real boundary condition at infinity in (\ref{eq:BCInfty}). (\emph{b}) Comparison between the experiment and the simulation for the radial position of point P. The green diamonds indicate the unmodified experimental data. Once a time shift i.e, a virtual origin for times $t_0=t_{0exp}=0.03$ is introduced, numerics (blue circles) and experiments (red dots) are in excellent agreement. Deviations from the scaling are expected when $t\sim1$, because of geometrical and gravitational effects that break the self-similar scaling.}
    \label{fig:zprpScaling}
\end{figure}

It is however crucial to notice that the Laplace pressure term $\kappa\Weber^{-1}$ in equation (\ref{eq:BernoulliSS}) is self-similar. This is because the result of balancing the Laplace pressure term in the Bernoulli equation (\ref{eq:BernoulliSS}) with the inertial terms on the left hand side using the self-similar ansatz (\ref{eq:Self2})-(\ref{eq:Self3}) will give us $2\beta=\alpha$ and $\beta=1-\alpha$, which is solved by exactly the same exponents $\alpha=2/3$, $\beta=1/3$ that we have just found by matching to the far field boundary condition (\ref{eq:BCInfty}). It is this remarkable feature that guarantees the existence of self-similar solutions to the system (\ref{Laplace2DSS})-(\ref{FreeSS}) for every value of the Weber number \Weber. To check our theoretical predictions, figure \ref{fig:zprpScaling} represents the function $x_P(t)$ calculated using the single fluid potential flow numerical code finding that, as expected, $x_P(t)\propto t^{2/3}$ for $t\ll 1$. For times $t\sim O(1)$ the $t^{2/3}$ self-similar behavior breaks because the non self-similar terms in equations (\ref{Laplace2DSS})-(\ref{FreeSS}) are no longer negligible. The experimental data included in figure \ref{fig:zprpScaling}($b$) shows that the radial position of point P can be fitted with a function proportional to $(t-t_{0exp})^{2/3}$, with $t_{0exp}\sim 0.03$ the virtual origin of times, which accounts for physical effects not taken into account in the theoretical derivation and the numerical simulations.

Figure \ref{fig:zprpScaling}($a$) shows that, contrary to $x_P(t)$, the time evolution of the vertical coordinate of point P calculated numerically, $z_P(t)$, is not purely proportional to $t^{2/3}$. As will be shown in \S\,\ref{subsec:selfsimcorrection}, this is due to the non self-similar effect of the downward motion of the disc or, equivalently, the upward motion of the undisturbed free surface in the frame of reference moving with the disc.\\

\begin{figure}
    \centering
    \includegraphics[width=6cm]{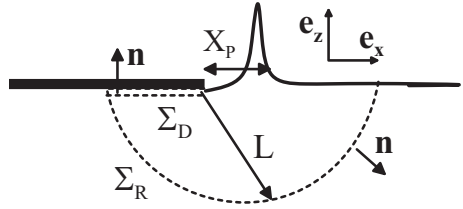}
    \caption{The surfaces $\Sigma_R$ (semicircle of radius $L\ll X_P$), $\Sigma_D$ (line of length $L$ located below the disc) and the free interface are the boundaries limiting the volume $\Omega(t)$ to which the integral balance of horizontal momentum is applied.
    }
    \label{fig:VolumenControl}
\end{figure}

An alternative way of understanding the $x_P\propto \tau^{2/3}$ scaling is obtained by making use of the fact that the boundary condition at infinity (\ref{BCInftySS}) implies that the disc acts as a constant `source' of momentum in time. Indeed, making use of the velocity field calculated from the potential (\ref{BCInftySS}), the variation per unit time of the horizontal momentum enclosed in the volume $\Omega$ defined in figure \ref{fig:VolumenControl} satisfies the following integral balance:
\begin{equation}
    \begin{split}
    &\mathbf{e}_x \cdot \frac{d}{dt}\int_{\Omega} \rho\mathbf{v}d\omega =\mathbf{e}_x \cdot \left( \int_\pi^{2\pi} \rho\mathbf{v} \mathbf{v}\cdot\mathbf{n}R_0 d\theta+\rho\,V_D\,\int_0^{R_0} \mathbf{v}\, dx-\int_\pi^{2\pi} p \mathbf{n} R_0 d\theta\right)=
    \\&
    -\rho\,A^2\,V^2_D\,R_D\int_\pi^{2\pi}\sin(\theta/2)
    \left(\sin(\theta/2)\cos\theta-\cos(\theta/2)\sin\theta\right)\,d\theta+\rho\,V_D\,\int_0^{R_0} \frac {\partial\,\phi}{\partial x}\,dx+
    \\&
    +\frac{1}{2}\rho\,A^2\,V^2_D\,R_D\int_\pi^{2\pi}\,\cos\theta\,d\theta=
    \rho\frac{\pi}{2}\,A^2\,V^2_D\,R_D+\rho\,V_D\,\int_0^{R_0} \frac{\partial\phi}{\partial x}\, dx=\rho\frac{\pi}{2}A^2\,V^2_D\,R_D+
    \\&
    +O(\rho\,V^2_D\,R_D(R_0/R_D)^{1/2})\, ,
    \label{Momentum}
    \end{split}
\end{equation}
where $\mathbf{n}$ is the outward normal vector to the surface $\Sigma_R$ defined in figure \ref{fig:VolumenControl}, $(\mathbf{e}_x,\mathbf{e}_z)$ are used to indicate the unit vectors along the coordinates $(x,z)$ and use of the steady Euler-Bernoulli equation namely, $p+1/2\rho\,\mathbf{v}\cdot\mathbf{v}=C$, has been made to express the pressure field $p$ over the surface $\Sigma_R$ of figure \ref{fig:VolumenControl} as a function of the velocity field. Notice that the contribution of the dimensionless flux of momentum through the surface $\Sigma_D$ defined in figure \ref{fig:VolumenControl} is of the order of $(R_0/R_D)^{1/2}\ll1$ and, therefore, can be neglected during the initial instants of the splash.

The result in equation (\ref{Momentum}) expresses that the time variation of the horizontal momentum contained in the volume $\Omega(t)$ is constant. We could have used this argument to deduce that there exists a self-similar solution for short times after impact. {Indeed, the characteristic length scale of the spatial region where the flow is time-dependent is $\sim X_P(T)$ and since there is not a characteristic time scale either, dimensional analysis indicates that velocities within this part of the splash region should scale as $\mathbf{v}\propto V_D\,x_P/\tau$. Therefore,}
\begin{equation}
    \mathbf{e_x} \cdot \frac{d}{dt}\int_{\Omega} \rho\mathbf{v}d\omega\propto \rho V^2_D R_D \frac{x_P^3}{\tau^2}\simeq \rho\,V^2_D\,R_D\frac{\pi}{2}A^2\rightarrow x_P\propto \tau^{2/3}\,, \label{selfsim}
\end{equation}
{where it has been taken into account that the area in the two-dimensional region where the flow is unsteady in the sketch of figure \ref{fig:VolumenControl} is
$\int_\Omega d\omega\sim R^2_D x^2_P$, concluding our alternative derivation of the scaling law for $x_P$.}
\bigskip

Clearly, the solution of the system (\ref{Laplace2DSS})-(\ref{FreeSS}) needs to be found numerically. It is, however, easier to solve the Laplace equation subjected to the unsteady boundary conditions given by (\ref{Bernoulli})-(\ref{Free}) and then express the solution in terms of the variables $\chi$, $\eta$ and $F$ defined in equations (\ref{eq:Self2})-(\ref{eq:Self3}), i.e., by just using the boundary integral method.

\begin{figure}
    \centering
    \includegraphics[width=9cm]{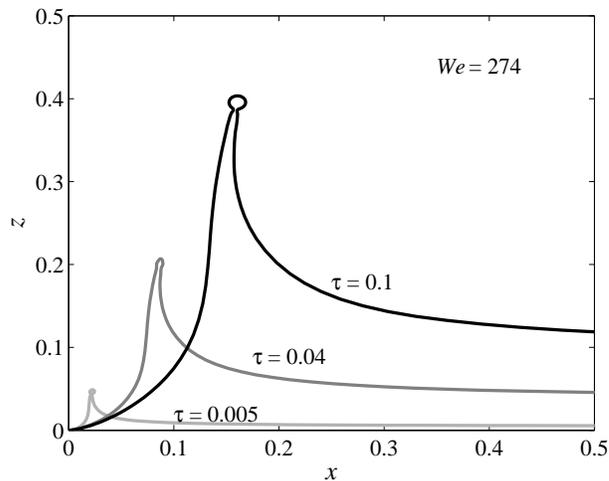}
    \caption{Time evolution of the splash region for a given value of the Weber number. The origin of the coordinate $z$ is located at the disc position. Notice that the height of the splash wave is much larger than the depth at which the disc is located. Considering a disc with a radius $R_D=20~\mathrm{mm}$ and impacting at a speed $V_D=1~\mathrm{m/s}$, the dimensionless times shown in this figure, $\tau=5\times 10^{-3}$, $\tau=4\times 10^{-2}$ and $\tau=10^{-1}$, correspond to $T=0.1~\mathrm{ms}$, $T=0.8~\mathrm{ms}$ and $T=2~\mathrm{ms}$ respectively.}
    \label{fig:unscaled_splash}
\end{figure}

\begin{figure}
    \centering
    \includegraphics[width=12cm]{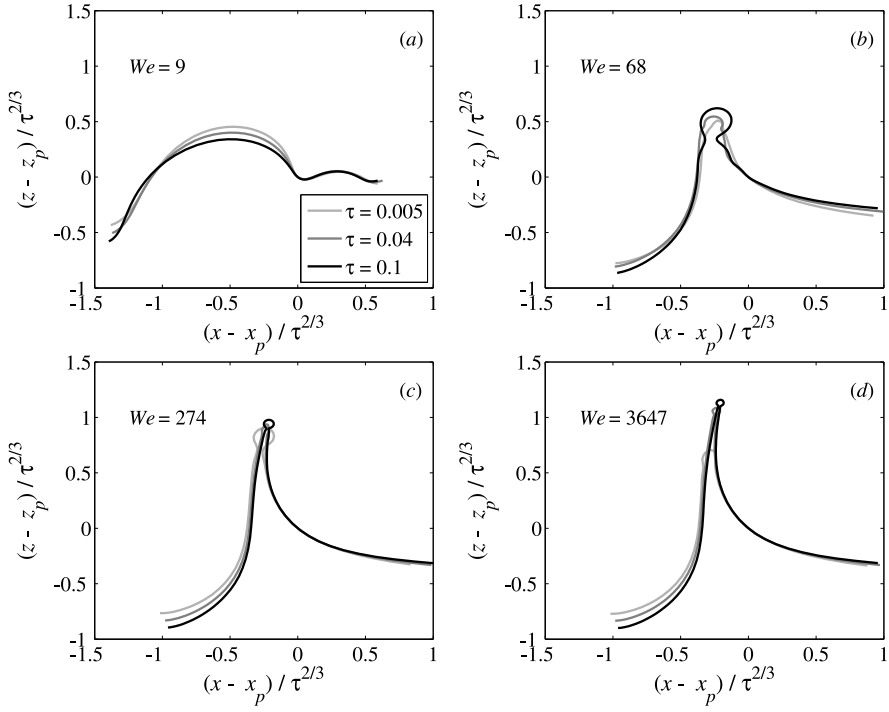}
    \caption{Shape of the splash with all distances rescaled by $\tau^{2/3}$ for four different values of the Weber number \Weber. Each plot contains three instances in time ($\tau=0.005$, $\tau=0.04$, $\tau=0.1$). The shapes have first been translated such that point P is in the origin of the plots, in order to better show the collapse of the different shapes.
    }
    \label{fig:scaled_splashes}
\end{figure}

\begin{figure}
    \centering
    \includegraphics[width=9cm]{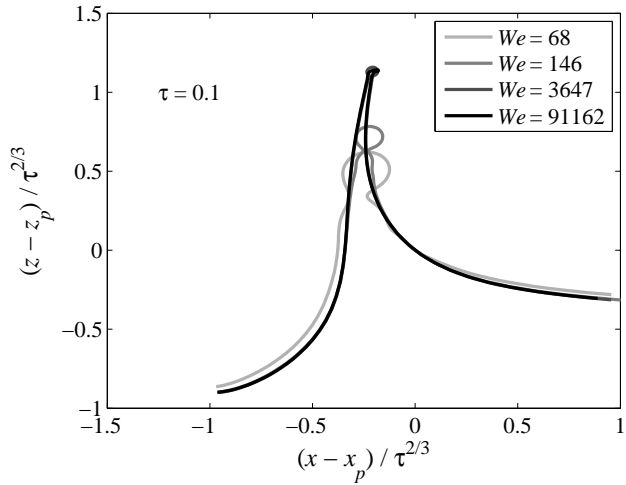}
    \caption{Comparison of the self-similar shape of the splash (i.e., with all length scales rescaled by $\tau^{2/3}$) for four different values of the Weber number. Differences in the rescaled shapes are only observed for the lower values of the Weber number, see for example the shape for $\Weber=68$ and $\Weber=9$ in figure~\ref{fig:scaled_splashes}. For high Weber numbers ($\Weber\gtrsim100$) the rescaled shape hardly changes as a function of \Weber.}
    \label{fig:selfsimprofiles}
\end{figure}

Figure~\ref{fig:unscaled_splash} shows the solution from the boundary integral method for one specific Weber number at three instances in time, which all are in the regime $\tau\ll1$ where we expect to find self similar solutions. In figure~\ref{fig:scaled_splashes}(\emph{c}) we have rescaled the solution of figure~\ref{fig:unscaled_splash}, according to (\ref{eq:Self2}). Inspecting all shapes in figure~\ref{fig:scaled_splashes}(\emph{a-d}), we see that indeed we find self similar solutions for a large range of Weber numbers, that only depend on \Weber. Figure~\ref{fig:selfsimprofiles} shows that for $\Weber\rightarrow\infty$ the solution becomes independent of \Weber, confirming the results of \cite{Iafrati2004}.

\subsection{Correction for non self-similar terms}
\label{subsec:selfsimcorrection}
\begin{figure}
    \centering
    \includegraphics[width=8cm]{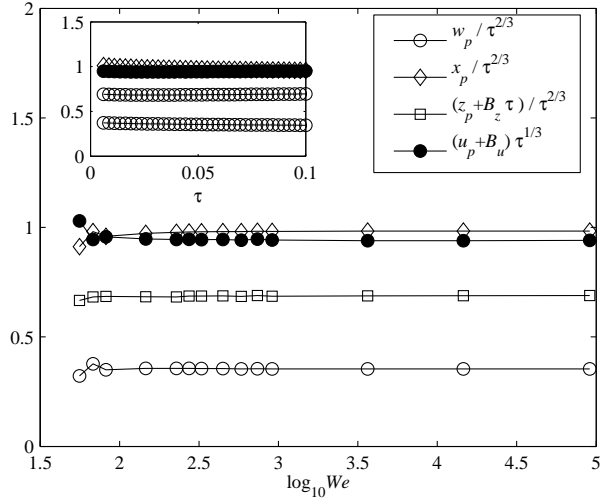}
    \caption{
    The positions $x_p$ and $z_p$, the splash width $w_p$, and the vertical velocity $u_p$, for different values of the Weber number.
    Clearly, all the rescaled values are independent of the Weber number for $\Weber\gtrsim100$, consistent with the fact that the self-similar shape of the splash becomes independent of $\Weber$ in this regime (see also figure~\ref{fig:selfsimprofiles}). The inset shows length scales and velocities in the splash rescaled by $\tau^{2/3}$ and $\tau^{-1/3}$ respectively. After proper rescaling, all the plotted values become independent of time, which indicates the self-similar behavior.
    Due to the boundary condition at the disc, a constant $B_u$ is added to the vertical velocity $u_p$ and a term $B_z$ times $\tau$ is added to the vertical position $z_p$ of point P (see text). All data in the inset correspond to a simulation at $\Weber=274$.}
    \label{fig:slope_and_tip}
\end{figure}

From figure~\ref{fig:zprpScaling} it is clear that the vertical position $z_p$ of point P does not follow the scaling $\tau^{2/3}$ strictly, except in the limit $\tau\rightarrow0$. This difference is caused by the downward motion of the disc (or, equivalently, the upward motion of the undisturbed free surface, since the problem (\ref{Laplace2DSS}-\ref{FreeSS}) was formulated in the frame of reference comoving with the disc), which introduces a non self-similar term in the far field boundary condition of equation~(\ref{FreeSS}). Nevertheless, we can compensate for the effect of the downward motion of the disc by introducing a constant $B_z$ such that
\begin{equation}
    (z_p+B_z\tau) \propto \tau^{2/3}. \label{Ecbz}
\end{equation}
A similar correction is needed for the vertical velocity $u_p$ at point P:
\begin{equation}
    (u_p+B_u) \propto \tau^{-1/3}.\label{Ecbu}
\end{equation}
The inset in figure ~\ref{fig:slope_and_tip} shows the position, velocity, and width of the splash at point P. All values become independent of time after the proper rescaling and figure~\ref{fig:slope_and_tip} shows the scaling of lengths (\emph{a}) and velocities (\emph{b}) for a wide range of Weber numbers. As expected from the previous analysis, the rescaled values become independent of \Weber\ for large Weber numbers. The same holds for the correction terms: $B_z=0.56\pm0.03$ and $B_u=0.59\pm0.01$ defined in equations (\ref{Ecbz})-(\ref{Ecbu}) for $\Weber\gtrsim50$.

\section{Crown breakup transition}
\label{sec:splashTransition}

We have provided a scaling for the shape of the splash, and have shown that the splash possesses a self-similar shape for every value of the Weber number. We will now have a closer look at the breakup of the splash into a crown. With the crown breakup, we are referring to the ejection of droplets from the tip of the splash.

In order to determine when the deceleration is large enough to provoke the growth of perturbations, we define a local Bond number at the tip of the splash using the radius of curvature and the acceleration felt by fluid particles in this region of the wave, namely, $A_{tip}-g$ (see equation~(\ref{eq:localBond})), where
\begin{equation}
    A_{tip}=-\frac{d\,U_{tip}}{d\,T}\, ,
\end{equation}
with  $-d\,U_{tip}/d\,T>0$. However, the determination of the local Bond number at the tip of the splash wave is not an easy task. Indeed, from the numerical simulations $\Bond_{tip}$ can be obtained very accurately, but only in an indirect manner for most Weber numbers. The reason for this is that the tip of the splash is unstable in the numerical simulations for $\Weber\gtrsim 30$, as is explained in more detail later in this section. Alternatively, the local Bond number at the tip can be determined directly from the experiments, but with a relatively large uncertainty. Both the experimental and numerical method of determining the local Bond number will be explained now.

\begin{figure}
\centering
    \includegraphics[width=\textwidth]{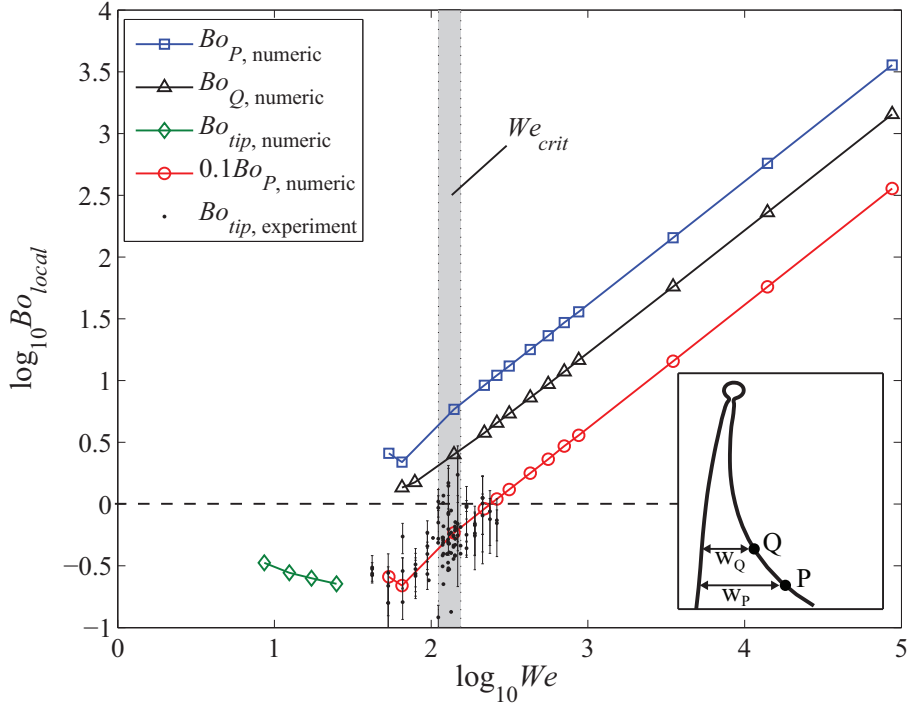}
    \caption{The local Bond number measured in different ways as a function of \Weber. The blue squares are the local Bond numbers determined in the simulations at point P ($\Bond_P$), the black triangles correspond to the local Bond number at the point Q ($\Bond_Q$) of the inset and red circles are these same values of $Bo_P$ multiplied by $0.1$. This multiplication factor has been found by matching to the experimental values of the local Bond number at the tip (black dots). The green diamonds show the local Bond numbers measured at the tip of the splash in the BI simulations, which is only possible for $\Weber\lesssim30$ (see text). The shaded area indicates the range of experimental conditions in which we observe the crown breakup transition. Note that this transition occurs when the local Bond number is of order unity (dashed horizontal line).
    }
    \label{fig:localBond}
\end{figure}

We obtain experimental values for $\Bond_{tip}$ by tracing the tip of the splash, within a time interval $T_i$ with a duration of typically $3~\mathrm{ms}$, in high-speed movies taken at 5400 frames per second. A second-order fit to the position data versus time gives us the mean deceleration of the tip $A_{tip}$. We determine the radius of the rim $R_C$ by measuring it graphically at the beginning and at the end of $T_i$, giving us a minimum and maximum value for $R_C$ within the time interval. The experimental values for $\Bond_{tip}$ are shown in figure~\ref{fig:localBond} in black dots. The error bars are obtained by using the minimum and maximum of $R_C$, because the radius of curvature is the dominant source of uncertainty, due to the squared appearance of $R_C$ in the definition of the Bond number (\ref{eq:localBond}).

\begin{figure}
\centering
    \includegraphics[width=\textwidth]{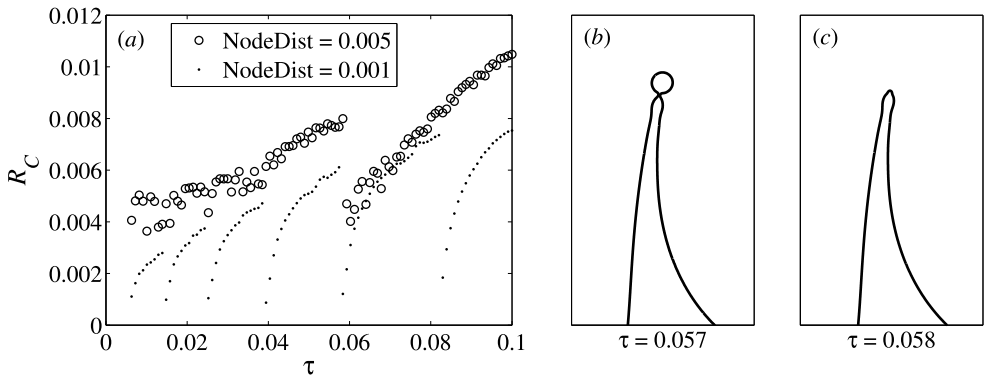}
    \caption{Influence of the node density on the tip of the splash, for $\Weber=274$. (\emph{a}) The radius of curvature $R_C$ at the tip has a minimum value comparable to the minimum distance between the nodes. The radius of curvature increases as the cylindrical rim grows in size, until the rim pinches off, and the radius of curvature starts again at its minimum value. This process is sensitive to the node density, but has no influence on the solution away from the tip, and is not present when \Weber\ is small enough. (\emph{b}) The cylindrical rim at its maximum size before it pinches off, for a minimum node distance of $0.001$. (\emph{c}) The same simulation as (\emph{b}), just after the rim has pinched off. The rim is removed from the simulation directly after the pinch-off because it does not represent the physical situation (see text).
    }
    \label{fig:NodeDensity}
\end{figure}
In the boundary integral simulations, the radius of curvature and deceleration can be determined accurately at any moment of time. The tip of the splash, which is the region in which we are interested, is however only stable in a limited range of low Weber numbers. For higher Weber numbers ($\Weber\gtrsim30$) a neck is formed below the rim, leading to its pinch-off from the rest of the splash wave (see figure~\ref{fig:NodeDensity}(\emph{b}-\emph{c}))\footnote{This pinched-off donut-shaped volume of fluid is subsequently removed from the simulation to prevent it from crossing the free surface in some point which would cause the code to crash.}. As soon as the pinch-off has occurred, a new neck forms which pinches off a bit later. This induces a series of pinch-offs in the numerics, which are unphysical for a number of reasons. The first is that, due to axisymmetry, the droplet is tubular, much unlike the droplets that are generated in experiments. The second is that the details of the pinch-off are strongly dependent on the node density that is used in the BI simulations, as is shown in figure~\ref{fig:NodeDensity}, and therefore needs to be qualified as a numerical artifact {associated with neglecting viscous effects in the computations. Indeed, we analyzed the numerical simulations and realized that the driving force leading to the detachment of the toroidal drop from the rest of the liquid mass is precisely the Bernoulli suction effect caused by the liquid acceleration at the neck region}. The series of pinches influence both the length scale and the deceleration of the tip, such that it is not possible to determine the local Bond number at the tip of the splash for high Weber numbers in the simulations. Although the tip of the splash is unstable, we can show that the rest of the splash is not influenced by these numerical artifacts. For this reason, we determine the local Bond number at point P, $\Bond_P$, using the width $W_p$ (see figure~\ref{fig:Esquema}) as the length scale and calculate the deceleration at t he same point, namely,
\begin{equation}
    \Bond_P = \frac{\rho}{\sigma}\left(-\frac{dU_P}{dT}-g\right)W^2_P.\label{BoP}
\end{equation}
In figure~\ref{fig:localBond}, the different methods of measuring the local Bond number are put together. Using the lower reference point P on the splash shows that for $\Weber\gtrsim100$, the local Bond number is proportional to \Weber\ (blue squares, figure~\ref{fig:localBond}), and we find that it is constant in time. Comparing $\Bond_P$ with the experimental values of $\Bond_{tip}$ however, shows that the local Bond number is overestimated if we use $\Bond_P$. But calculating the local Bond number in the numerical simulations at different positions on the splash wave results only in a vertical shift of the points in figure~\ref{fig:localBond}. Indeed, figure \ref{fig:localBond} shows that the value of the local Bond number evaluated at the point Q on the splash wave where the slope is -2 (thus, point Q is closer to the edge of the rim than point P), is proportional to $Bo_P$ defined in equation (\ref{BoP}). This result is a consequence of the fact shown in figure \ref{fig:slope_and_tip} that the dependence of the lengths and velocities in the splash wave are not dependent on the Weber number for sufficiently large values of this control parameter. This evidence indicates that we can calculate the local Bond number at different positions by simply multiplying $\Bond_P$ by a constant. For $\Bond_{tip}$, we determine this constant to be $\sim0.1$ using the experimentally measured local Bond number at the tip of the splash. Also notice in figure \ref{fig:localBond} that the value for the constant $\sim 0.1$ is also consistent with the values of $Bo_{tip}$ numerically calculated for low values of the impact Weber number. The local Bond number at the tip that we have deduced from $\Bond_P$ is shown in figure~\ref{fig:localBond} with red circles. Clearly, the transition to breakup into a crown occurs when the local Bond number at the tip of the splash is of order unity.

The proportionality of the local Bond number with the Weber number and its independence of time is not unexpected, because it results from the self-similarity solution that becomes independent of \Weber\ for large Weber numbers, as shown in figure \ref{fig:slope_and_tip}. Using the self-similar scaling for distances, velocities and accelerations, we write the deceleration and typical length scale in dimensional form as:
\begin{equation}
    A_{tip} = \frac{V_D^2}{R_D} C_a \tau^{-4/3}
    \label{eq:accDimensional}
\end{equation}
and
\begin{equation}
    R_C = R_D C_R \tau^{2/3} \, ,
    \label{eq:lengthDimenional}
\end{equation}
where $C_a$ and $C_R$ are dimensionless constants, independent of time and Weber number for $\Weber\gg 1$. Substituting (\ref{eq:accDimensional}) and (\ref{eq:lengthDimenional}) in (\ref{eq:localBond}), and using the fact that $A_{tip}\gg g$ for large values of \Weber\ gives:
\begin{equation}
    \Bond_{tip} \simeq C_a C_R^2 \frac{\rho V_D^2R_D}{\sigma} = C_a C_R^2 \Weber \, ,
\end{equation}
clearly showing that the local Bond number is independent of time and proportional to \Weber.

Using the proportionality $\Bond_{tip}\propto\Weber$ and the crown breakup condition that $\Bond_{tip}$ is of order unity around the transition, we can define a condition for the crown breakup transition based on \Weber\ that does not involve the local Bond number. Such a condition is preferable, because whereas the local Bond number is difficult to determine, the Weber number is directly given by the experimental conditions. There indeed exists one critical Weber number $\Weber_{crit}$ above which we always observe breakup of the splash into a crown, as can be seen in table \ref{tab:We_crit}. The table shows the values of \Weber, \Froude\ and \Rey\ at the crown breakup transition\footnote{Clearly, \Weber, \Froude\ and \Rey\ can be related after calculating the disc velocity $V_D$ using the disc radius $R_D$ and either of the three dimensionless numbers in table \ref{tab:We_crit}.}. Note that only the Weber number always has approximately the same value at the transition, showing that the Weber number is indeed the relevant parameter to indicate the transition to crown breakup and droplet ejection.

\begin{table}
  \begin{center}
  \begin{tabular}{cccc}
      $R_D~\mathrm{(mm)}$  & $\Weber_{crit}$ & \Froude\ & \Rey\ \\[3pt]
       15 & $145\pm5$  & $4.77$ & $1.26\cdot10^4$\\
       20 & $135\pm19$ & $2.51$ & $1.40\cdot10^4$\\
       25 & $134\pm11$ & $1.59$ & $1.56\cdot10^4$\\
       30 & $142\pm19$ & $1.18$ & $1.77\cdot10^4$\\
  \end{tabular}
  \caption{Transition to crown breakup of the splash for different disc radii. The value of the critical Weber number does not differ appreciably with the size of the disc, where the Froude number and the Reynolds number at the transition show a clear dependence on $R_D$.}
  \label{tab:We_crit}
  \end{center}
\end{table}


\section{Air cushion effect: force on the disc and drop size}
\label{sec:air}

The previous theoretical derivation of the self-similar solution predicts decelerations within the splash wave proportional to $\propto t^{-4/3}$, with $t$ the dimensionless time after impact. The huge decelerations expected from the analysis for $t\ll 1$ are not observed in a real experiment since there are other physical effects, not taken into account in the previous theoretical description, that break the self similar structure of the flow field during a short time interval after impact. This is clearly seen in figure \ref{fig:Xs}, where it is depicted that the splash wave initially propagates radially outwards at a velocity \emph{constant} in time, and not at a diverging velocity proportional to $\propto t^{-1/3}$, as 
is predicted by the self-similar solution. The origin of the cutoff time or, equivalently, the origin of the cutoff length below which the self similar solution is no longer valid, could in principle lie in any of the following three sources: (i) the fact that the disc edge is rounded, (ii) the effect of liquid viscosity, and (iii) the air layer between the impacting disk and the liquid.

First we assess the fact that the disc edge is not perfectly sharp, but possesses a finite curvature. Since, in view of figure \ref{fig:zprpScaling} the self-similar solution is only valid for $t>t_{0exp}\sim 0.03$, we can estimate the cut-off length $X_0=x_0R_D\sim R_Dt_{0exp}^{2/3}\sim 1.9~\mathrm{mm}$. This is comparable with the disc thickness and clearly much larger than the length associated with the finite radius of curvature at the disc edge which we estimate to be $\approx 50\,\,\mu\text{m}$. To evaluate the effect of liquid viscosity we estimate the viscous cutoff time, $t_{0v}$, whose magnitude is determined by equating the boundary layer thickness, $\delta$, to the width of the splash region at the instant of time when the self similar solution begins to be valid, namely,
\begin{equation}
    \delta\sim \sqrt{\frac{\mu R_D\,t_{0v}}{\rho\,V_D}}\sim
    R_D\,t^{2/3}_{0v}\rightarrow t^{1/6}_{0v}\sim Re^{-1/2}_D\rightarrow
    t_{0v}\sim Re^{-3}_D\sim 10^{-10}\, . \label{t0v}
\end{equation}
Clearly, this value is much smaller than $t_{0exp}$ and viscosity can therefore not account for the observed cut-off time.

\begin{figure}
    \centering
    \includegraphics[width=12cm]{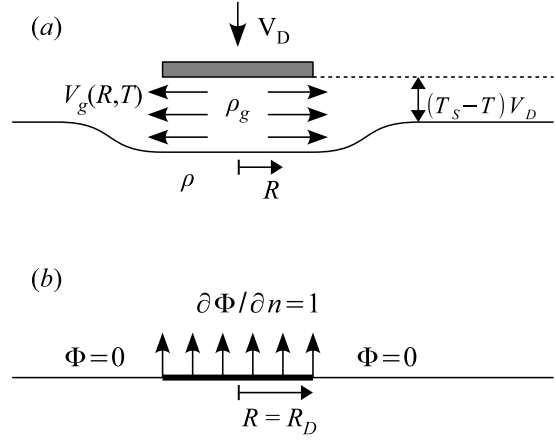}
    \caption{a) The downward motion of the disc forces the gas to flow radially outwards. The overpressure needed to accelerate the gas deforms the interface. b) The linearity of the Laplace equation permits to express the liquid potential at $R=0$ for a generic value of $d\,H/d\,T$ as $\Phi(R=0,T)=-\alpha R_D\,d\,H/dT$, with $\alpha R_D V_D$ the value of the potential at $R=0$ obtained from the solution of the Laplace equation subjected to the boundary conditions depicted in this figure. Here, $\partial \Phi/\partial n$ indicates the normal velocity at the interface.
 }
    \label{fig:gassketch}
\end{figure}

The remaining physical effect is the finiteness of air density, which is known to be responsible for the so called \emph{air cushion effect}, analysed by \cite{Howison,Wilson} for the case of two dimensional geometries. Figure \ref{fig:gassketch}a illustrates the disc approaching the free interface forcing the air between disc and liquid surface 
to flow radially outwards. Using continuity and neglecting compressibility effects, the radial velocity field of the gas, $V_g(R,T)$, can be expressed as a function of (dimensional) time $T$ and the vertical deformation $H_g(T)$ of the free surface below the disc as
\begin{equation}
\begin{split}
& -\pi\,R^2\left(V_D-\frac{d\,H_g}{d\,T}\right)+2\pi\,R\,\left[H_g+V_D\left(T_s-T\right)\right]\,V_g=0\quad\Rightarrow\\& 
    V_g=\frac{R}{2}\frac{\left(V_D-d\,H_g/dT\right)}{H_g+V_D(T_s-T)}\,
    ,\label{Eq0air}
    \end{split}
\end{equation}
where we assume that $H_g$ in the region below the disc does not depend on $R$. In (\ref{Eq0air}), $T_s$ is a constant which will be arbitrarily fixed to $T_s=R_D/V_D$ since its precise value possesses a negligible influence on $H_g(T)$.
Using the expression for the flow field given in (\ref{Eq0air}), the gas pressure dependence on $R$ and $T$ can be determined integrating the momentum equation
\begin{equation}
\rho_g\frac{\partial V_g}{\partial T}+\frac{\partial}{\partial\,R}\left(\rho_g\frac{V^2_g}{2}+P_g\right)=0\, ,
\end{equation}
which yields the following expression for $P_g(R,T)$:
\begin{equation}
    P_g-P_{a}=\frac{3}{8}\rho_g\left(R^2_D-R^2\right) \left(\frac{
    V_D-d\,H_g/dT}{H_g+V_D(T_s-T)}\right)^2-\rho_g\frac{R^2_D-R^2}{4}\frac{1}{H_g+V_D(T_s-T)}\,\frac{d^2\,H_g}{d\,T^2}\, .\label{Pair}
\end{equation}

The desired equation for $H_g(T)$ is determined from the Euler-Bernoulli equation particularized at $R=0$, namely,
\begin{equation}
\rho\frac{\partial\Phi}{\partial\,T}(R=0,T)+\frac{1}{2}\rho \left(\frac{d\,H_g}{d\,T}\right)^2+P_g(R=0)=P_a\, ,\label{Bernair}
\end{equation}
where $P_a$ indicates the atmospheric pressure. Indeed, assuming that the normal velocity in the liquid below the disc does not depend on $R$, the potential $\Phi$ at $R=0$ can be expressed as a function of $H_g$ since, by virtue of the linearity of the Laplace equation, $\Phi(R=0)=-\alpha\,R_D\,d\,H_g/d\,T$, with $\alpha=2/\pi$ the value of the potential at $R=0$ corresponding to the solution of $\nabla^2\Phi=0$ subjected to the boundary conditions sketched in figure \ref{fig:gassketch}b (see \cite{Iafrati-Korobkin11}). Finally, the substitution of equation (\ref{Pair}) into (\ref{Bernair}) and subsequent non-dimensionalization provides the following
differential equation for $h=H_g/R_D$:
\begin{equation}
-\left(\alpha+\frac{1}{4}\frac{\rho_g}{\rho}\frac{1}{t_s-t+h}\right)\frac{d^2\,h}{d\,t^2}+\frac{1}{2}\left(\frac{d\,h}{d\,t}\right)^2+\frac{3}{8}\frac{\rho_g}{\rho} \left(\frac{1-d\,h/dt}{h+t_s-t}\right)^2=0\, , \label{Ecdifgas}
\end{equation}
with $t = TV_D/R_D$ and similarly for $t_s$.

\begin{figure}
    \centering
    \includegraphics[width=14cm]{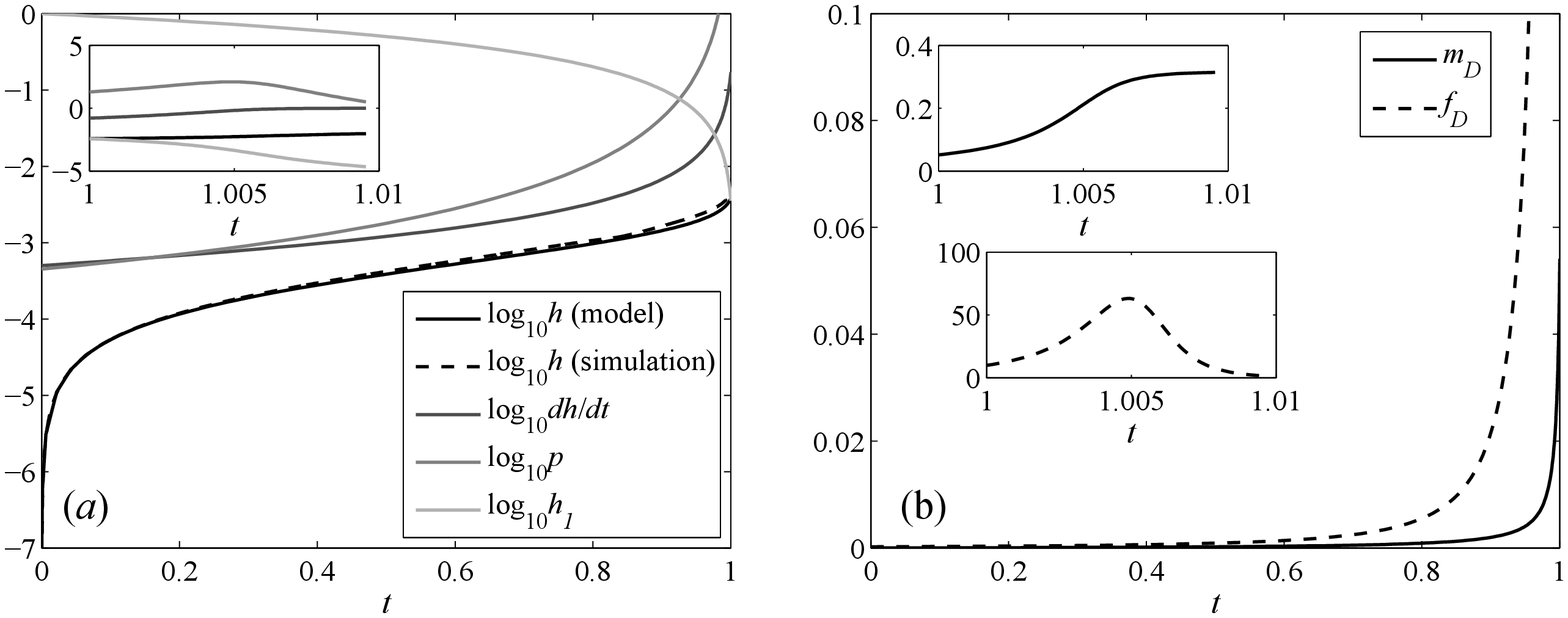}
    \caption{($a$) Using continuous lines, we plot the time evolutions of the gas layer depth $h$, the liquid velocity at $r=0$, $dh/dt$, gas pressure at $r=0$, $p_g$, and the distance from the disc bottom to the liquid interface, $h_1=t_s-t+h$ for $\rho_g/\rho=1.2\times 10^{-3}$ and $\alpha=2/\pi$. The inset shows the time evolution of the different variables for instants of time right after the disc bottom is below the undisturbed free interface. The discontinuous line represents the function $h(t)$ obtained from the two-fluid potential flow code, which is almost perfectly reproduced by the integration of equation (\ref{Ecdifgas}). ($b$) Time evolution of the functions representing the dimensionless force on the disc $f_D$ (dashed lines) and $m_D$ (continuous lines) defined, respectively, in equations (\ref{EcfD})-(\ref{Momentoair}). The insets show that due to the effect of air the force is not infinite any more, but reaches a well-defined maximum slightly after the disc bottom reaches the level of the unperturbed free interface and that the maximum value of the dimensionless change in momentum, defined in (\ref{Momentoair}), is $m_D(t=1.01)\simeq 0.315\sim O(1)$.}
    \label{fig:resultsgas1}
\end{figure}

{Let us point out that the above analysis rests on the assumption that viscous effects in the gas are unimportant, and this is indeed the case since the characteristic length $L_{e}$ in which viscosity transforms a plug flow into a parabolic Poiseuille velocity profile is just the entrance length in pipe flows, which is such that}
\begin{equation}
    \frac{L_e}{R_D}\sim \frac{\rho_g V_{g} H_1}{\mu_g}\frac{H_1}{R_D}=Re_g\frac{H_1}{R_D}\, ,\label{Regas}
\end{equation}
{where $V_g$ is given in equation (\ref{Eq0air}), $Re_g=\rho_g V_D R_D/\mu_g$, $H_1=V_D(T_S-T)+H_g$ and where use of the continuity equation (\ref{Eq0air}) has been made. The estimation in equation (\ref{Regas}), which is valid while $V_D-dH_g/dT\sim O(V_D)$, suggests to neglect viscous effects when $H_1/R_D\gtrsim 10 Re_g^{-1}$. In our experiments $Re_g\sim O(10^3)$ and, therefore, we can assume that the radial velocity profile in the gas is uniform while $H_1/R_D\gtrsim 10^{-2}$. The estimation in equation (\ref{Regas}) also suggests that viscous effects in the gas should be included in the analytical description if $H_1<10^{-2}\,R_D$. However, when this condition is fulfilled, the liquid normal velocities are already close to that of the impactor, as shown in figure \ref{fig:resultsgas1}. Thus, the gas overpressures associated to gas inertia set the liquid into motion before gas viscosity needs to be retained into the analytical description of the air cushion effect described above}.\\

\begin{figure}
    \centering
    \includegraphics[width=8cm]{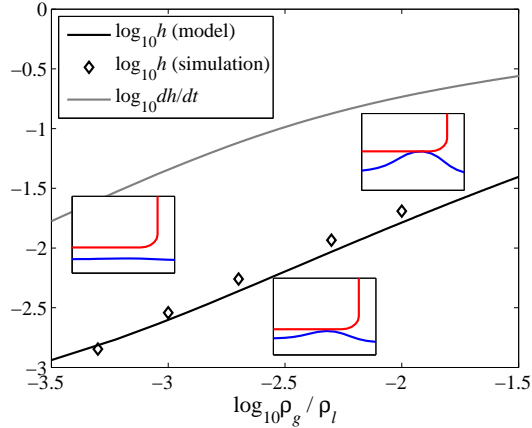}
    \caption{Continuous lines indicate the dependence of both $h$ and $d\,h/dt$ on
    the density ratio $\rho_g/\rho$ at $t=0.99$ predicted by the integration of equation (\ref{Ecdifgas}). Diamonds indicate the numerical values of the air layer thickness at $r=0$ when the first contact between the liquid and the disc bottom occurs, which occurs near the disc edge, as shown by the different insets. Numerical simulations reveal that, except close to
    the disc edge, the air layer depth is mostly uniform.}
    \label{fig:resultsgas2}
\end{figure}

The integration of equation (\ref{Ecdifgas}), subjected to appropriate initial conditions, provides with the time evolution for $h$, $dh/dt$, the dimensionless gas pressure $p(r=0)=P_g/(\rho\,V^2_D)$ and the height of the gas layer $h_1=t_s-t+h$ shown in figures \ref{fig:resultsgas1}-\ref{fig:resultsgas2}. It is depicted in figure \ref{fig:resultsgas1}(a) that the interface is only appreciably accelerated for $t\simeq 1$ i.e., when the disc is sufficiently close to the undisturbed interface. Moreover, only for times such that $t\gtrsim t_{end}$ with $t_{end}=1.01$, namely, when the disc position is only slightly below the level of the undisturbed interface, $d\,h/dt\simeq 1$ and $h_1\simeq 0$. Therefore, the liquid normal velocities below the disc are smaller than the disc impact velocity while $t<t_{end}$ and equal to the disc velocity for $t>t_{end}$. This fact implies that the speed at which the splash wave propagates must be, during a time interval $t-t_s\lesssim 0.01$, smaller than the one predicted by the previous potential flow numerical simulations where the effect of air was neglected. This is just what figure \ref{fig:Exp} shows: the splash wave calculated numerically needs 0.6 ms less than the real wave to reach the wave radial position at $T=0.2ms$ ($\tau=0.1$); from that spatial position onwards, the wave speed calculated numerically is identical to that measured experimentally. We can conclude that the air layer below the disc breaks the self similar structure of the splash wave during $t_{end}-t_s\simeq 0.01$, a time interval very similar to the magnitude of the virtual origin in times, $t_{0exp}=0.03$, deduced in figure \ref{fig:zprpScaling}.

{Let us point out that the analysis of the experimental data for different impact velocities and disc radii reveals a dispersion in the values of $t_{exp}$ too large to find a clear trend. Nevertheless, since all the measured values are of the order of $0.01$, together with the fact that radial position of the splash waves collapse onto a single curve after rescaling with $R_D$ and $V_D$ (see figure~\ref{fig:Xs}$b$), suggest that $t_{exp}$ is independent of \Weber, in accordance with our theory.}

To check the validity of the model presented above, we have carried out two-fluid potential flow numerical simulations using the numerical code detailed in \cite{GordilloEtAl_PoF_2007,GekleGordillo2011}. Both the density ratio $\rho_g/\rho$ and the impact Weber number have been varied in wide ranges of values. Figure \ref{fig:resultsgas1} shows that the time evolution of the dimensionless height $h$ is almost perfectly reproduced by our theory. The shape of the entrapped bubble calculated numerically for different values of the density ratio, depicted in figure \ref{fig:resultsgas2}, shows that the interface accelerates beneath the disc uniformly downwards, except in a very localized region near the disc edge, where the interface accelerates upwards as a consequence of Bernoulli suction (or, equivalently, due to the growth of a Kelvin-Helmholtz instability), causing the entrapment of an air pocket. It is also shown in figure \ref{fig:resultsgas2} that the maximum height of the gas layer calculated numerically compares very favorably with the one predicted theoretically in spite of the fact that our model, which was built under the assumption that $h$ does not depend on $r$, is unable to capture the upwards acceleration of the free interface. Let us point out here that the air cushion effect described above shares many similarities with the bubble entrapment processes taking place either during the impact of drops against solid or liquid surfaces \cite{ThoroddsenDrop,MandreBrenner,Josserand} or in the collapse of bubbles \cite{PRL07,PoF08}. It should be mentioned here that, while the gas flow is well described by the lubrication approximation in the impact of drops, viscous effects are negligible in our case.

\begin{figure}
    \centering
    \includegraphics[width=14cm]{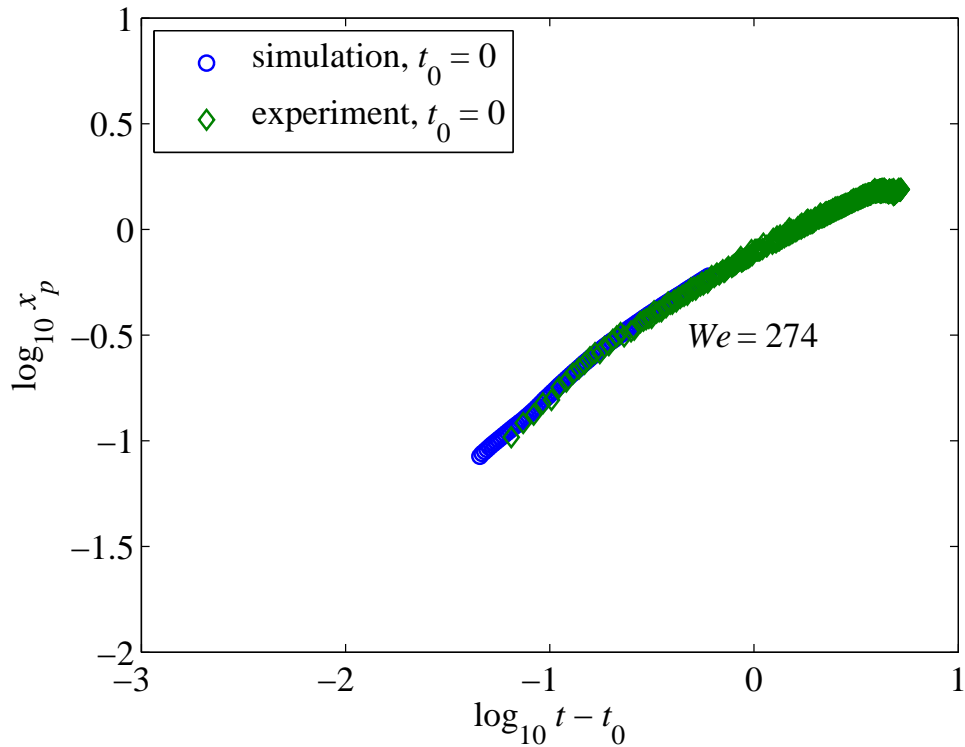}
    \caption{Comparison between the time evolution of the horizontal position $x_P$ of point P for $\Weber=274$ in (i) a single-fluid potential-flow numerical simulation in which the liquid normal velocities beneath the disc is provided by our theoretical prediction (function $dh/dt$ in figure \ref{fig:resultsgas1}a, blue circles) and (ii) the experiments (green diamonds). Note that the experimental data is the bare data of figure~\ref{fig:zprpScaling}b without the time shift. The conclusion is that the simulations based on our theoretical analysis excellently reproduces the experiments, with no adjustable constants.}
    \label{fig:resultsgas3}
\end{figure}

The critical role played by the thin layer of air located beneath the disc on the splash wave speed can be demonstrated in an even more convincing way. To this end, we have carried out single-fluid potential-flow simulations in which the effect of air is introduced by replacing the constant impact velocity boundary condition by the function $d h/d t$ represented graphically in figure \ref{fig:resultsgas1}a. Figure (\ref{fig:resultsgas3}) shows that this new type of numerical simulations, which indirectly retain the air cushion effect through its influence on the normal velocity beneath the disc, perfectly reproduces the radial propagation of the splash wave and, thus, avoids the need of using an artificial time shift to match simulations with experiments. {The good agreement between our analytical model and experimental results also confirms that the effect of gas viscosity on the splash wave formation is small when compared to that of gas inertia, as anticipated above.}

Once the critical role played by air in the impact process has been shown to be very well described by our theory, we determine the force on the disc, $F_D(T)$, by integration of the pressure field given by equation (\ref{Pair}), to give
  \begin{equation}
  F_D(T)=\rho\,\pi\,R^2_D\,V^2_D\,f_D(t)\, ,
  \end{equation}
  where the function
  \begin{equation}
  f_D(t)=\frac{1}{16}\frac{\rho_g}{\rho}\left[3\left(\frac{1-dh/dt}{h+t_s-t}\right)^2-\frac{2}{h+t_s-t}\frac{d^2\,h}{dt^2}\right] \label{EcfD}
  \end{equation}
  is represented graphically in figure \ref{fig:resultsgas1}(b) for the particular case of $\rho_g/\rho=1.2\times 10^{-3}$. Figure \ref{fig:resultsgas1}(b) shows the remarkable result that $f_D(t)$ reaches a maximum when the disc bottom is below the level of the unperturbed free surface but before the disc bottom touches the liquid. Let us point out here that the force on an impacting disc was already calculated by \cite{Iafrati2008,Iafrati-Korobkin11}, who neglected the effect of air in the analysis, a starting hypothesis leading to the force on the disc tend to infinite during the initial instants of the impact process. The fact that our theory predicts a maximum force on the disc, indicates that the air cushion effect described above softens the impact process, thus avoiding the diverging force deduced from the analysis when air is removed. Our theory also allows to determine the change in the momentum of a disc falling freely and impacting against a liquid free surface,
  \begin{equation}
  \Delta\,M_D(T)=\int_0^{T} F_D(T')\,d\,T'=\rho\,\pi\,R^3_D\,V_D\,m_D(t)\, ,\label{Momentoair}
  \end{equation}
  where the function $m_D(t)$ defined in equation (\ref{Momentoair}) is also represented graphically in figure \ref{fig:resultsgas1}(b). Since $m_D(t=1.01)\simeq 0.315$, the change in velocity $\Delta\,V_D$ of a free falling disc of density $\rho_s$ and thickness $H$ impacting normally onto water-air interface ($\rho_g/\rho=1.2\times 10^{-3}$) can be estimated as:
  \begin{equation}
  \rho_s\pi\,R^2_D\,H\,\rho_s\Delta V_D\simeq 0.315 \pi\,\rho\,V_D\,R^3_D\rightarrow \frac{\Delta\,V_D}{V_D}\simeq 0.315\frac{\rho}{\rho_S}\frac{R_D}{H}\, .\label{Momentoair2}
  \end{equation}
{The result expressed by equation (\ref{Momentoair2}) has been obtained under the assumption that the disc velocity is constant and, thus, it can only be strictly applied in the case in which the velocity of the solid does not appreciably vary during the impact process. That is, when the increment in the disc momentum given by equation (\ref{Momentoair}) is small when compared with the initial momentum. Nevertheless, equation (\ref{Momentoair2}) can be used for qualitative estimation purposes. Doing so, we can conclude that, since for most solids $1<\rho_s/\rho<20$, the air cushion effect could reduce the impacting velocity by an amount $\sim O(V_D)$ if $R_D/H>O(10)$.} Let us finally point out that the validity of our study is restricted to those situations in which the disc impact velocity is sufficiently low so as to neglect gas compressibility effects.

\subsection{Drop size}

\begin{figure}
\centering
    \includegraphics[width=14cm]{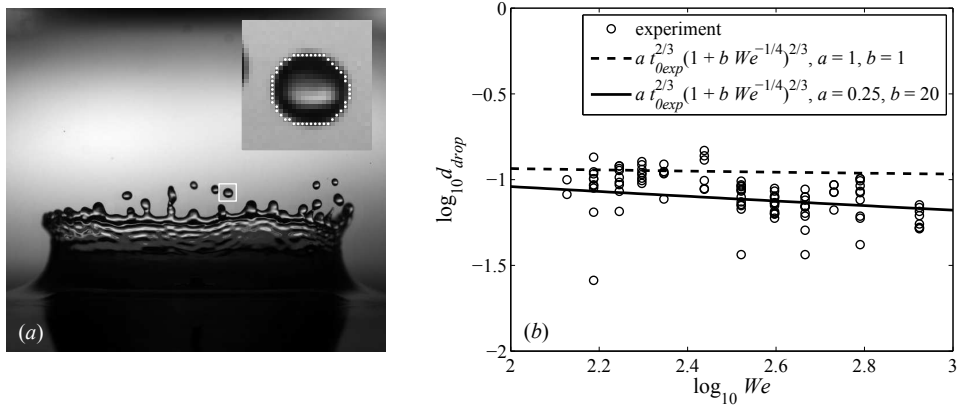}
    \caption{a) Snapshot showing the crown splash and how the edge finding routine identifies the ejected drops. Drop diameter is experimentally determined from the analysis of this type of images. b) This figure compares the diameter of the ejected drops with the theoretical prediction given by equation (\ref{dropsize}) for two different values of $b$.
    }
    \label{fig:dropDiameter}
\end{figure}

It was demonstrated above that, in the range of times in which the self-similar solution is valid (i.e., when $t_{0exp}\ll t\ll 1$), the local Bond number is constant in time. Indeed, during this time interval, distances and decelerations in the tip region are proportional to $R_C/R_D=r_{tip}\propto (t-t_{0exp})^{2/3}$ and $a_{tip}\propto (t-t_{0exp})^{-4/3}$ respectively and, thus,
\begin{equation}
    Bo_{local}=\frac{\rho\,V^2_D\,R_D}{\sigma} a_{tip} r^2_{tip}\propto We.
\end{equation}
Therefore, whenever $Bo_{local}>1$ or equivalently, if $We>We_{crit}$, it is expected that when the rim decelerates i.e., at any instant of time such that $t>t_{0exp}$, the rim develops a Rayleigh-Taylor instability with a initial length scale $d<r_{tip}$ given by the condition (see \cite{VillermauxBossa})
\begin{equation}
    \frac{\rho V_D^2\,R_D a_{max,tip} d^2}{\sigma}\simeq 1\rightarrow d\simeq We^{-1/2} a_{max,tip}^{-1/2} \, , \label{equationd}
\end{equation}
where it has been taken into account that the maximum deceleration at the tip of the splash wave is such that $a_{max,tip}\gg g R/V^2_D$. Equation (\ref{equationd}) expresses that the characteristic length scale of the fingers, $d\,R_D$, is set by the condition that their weigh per unit length, namely, $\rho\,V^2_D R_D\,a_{tip} d^2$, equals the surface tension force per unit length, $\sigma$. The maximum deceleration scales as $a_{max,tip}\propto t^{-4/3}_{0exp}\sim \left(t_{end}-t_s\right)^{-4/3}$ and, therefore, the initial wavelength of maximum growth rate is given by
\begin{equation}
    d\propto t^{2/3}_{0exp}\,We^{-1/2}\, ,\label{scaled}
\end{equation}
where use of equation (\ref{equationd}) has been made. These initial perturbations give rise to the formation of fingers of increasing width and length which break due to the action of surface tension forces in a characteristic capillary time given by
\begin{equation}
    t_{cap}=V_D T_{cap}/R_D=V_D/R_D\,\left(\rho R^3_D d^3/\sigma\right)^{1/2}=We^{1/2} d^{3/2}\propto We^{-1/4}\,t_{0exp}\, ,
\end{equation}
where use of equation (\ref{scaled}) has been made. Consequently, the breakup time will be given by $t_{break}\propto (t_{0exp}+t_{cap})$ and the dimensionless size of the drop formed, which is of the order of the width of the rim (see figure the experimental evidence in figure \ref{fig:dropDiameter}a), is given by
\begin{equation}
    \begin{split}
    &d_{drop}\sim w_{rim}\propto t_{break}^{2/3}\propto t^{2/3}_{0exp}\left(1+t_{cap}/t_{0exp}\right)^{2/3}\propto t^{2/3}_{0exp}\left(1+b\,We^{-1/4}\right)^{2/3},\label{dropsize}
    \end{split}
\end{equation}
with $b$ an adjustable constant. Figure~\ref{fig:dropDiameter}b shows a reasonable agreement between the experimental drop size measured from the analysis of the images of the type depicted in figure~\ref{fig:Exp0} and the drop diameter predicted by equation (\ref{dropsize}). Let us point out that figure~\ref{fig:dropDiameter}b reveals that the drop size is hardly dependent of the impact Weber number. This fact indicates that the characteristic length scale of the ejected drops is fixed by the air bubble entrapped underneath the disc, being this the physical mechanism underlaying the experimentally observed cut-off time, $t_{0exp}$.

{Let us point out that the broad distribution of drop sizes depicted in figure~\ref{fig:dropDiameter}b is most likely caused by small disturbances of different initial amplitudes distributed unevenly along the azimuthal direction. Indeed, a regular periodic structure of drop ejection along the rim is expected only when there is an equal amount of noise in all wavelengths because, only in that case, we would observe the growth of just the wavelength selected by the Rayleigh-Taylor type of instability, i.e, the one with the largest growth rate. Among the reasons that could lead to the initial amplitude of perturbations to depend on the wavelength and on the azimuthal direction are: a slight deviation from axisymmetry, since this will cause the radial component of the gas velocity to vary in the azimuthal direction, a small tilt of the impacting disc, geometrical imperfections on the disc itself or the presence of waves before the disc impacts the free surface.}

\section{Conclusions}
\label{sec:conclusions}

We have studied the formation of a splash wave and the transition to the ejection of droplets after the impact of a disc on a liquid by using boundary integral simulations, experiments and theoretical analysis. Only by combining these three methods, we have been able to analyze the full problem. Although all the different approaches used possess limitations either in accuracy, stability or mathematical formulation, each of them is able to reveal specific parts of the problem unaccessible by the other methods. Using the overlapping parts however, we accurately demonstrated the validity of each of the three different approaches used in this study.

By approximating the time just after disc impact ($t\ll1$) as a 2-dimensional potential flow problem and neglecting the effect of air, we have shown that there exist self-similar solutions of the second kind for any value of the Weber number. These self-similar solutions exist because the matching of the inertial terms in the unsteady Bernoulli equation to the far-field boundary condition for the potential gives the same scaling powers as the matching with the surface tension term. When the Weber number is increased to large values ($\gtrsim100$), the shape of the splash becomes independent of the Weber number. Both predictions are confirmed by boundary integral simulations, by rescaling the calculated shapes for a wide range of Weber numbers. We found the correct scaling for both the lengths ($\sim t^{2/3}$), and velocities ($\sim t^{-1/3}$) at different positions in the splash. In the experiments, the same scaling is found after introducing a small time-shift $t_{0exp}$ which we demonstrated is associated to the finiteness of gas density. Indeed, in spite of its large density contrast with water or the solid, the very thin air cushion entrapped between the impacting solid and the liquid reveals that gas critically affect the velocity of propagation of the splash wave as well as the force on the disc, $F_D$. The presence of air avoids the singularity in $F_D$ predicted when air effects are neglected (see \cite{Iafrati-Korobkin11}) and our theory predicts that the maximum of $F_D$ is reached \emph{before} the disc touches the liquid.

We have shown that the transition to droplet ejection (crown breakup transition) is caused by a Rayleigh-Taylor instability. We experimentally determined that the local Bond number based on the tip deceleration and the tip width, is of order unity at the splash transition. Theoretical analysis predict, and numerical simulations confirm that, $\Bond_{tip}$ depends linearly on the Weber number for $\Weber\gtrsim100$. From this linear dependence between $\Bond_{tip}$ and \Weber, we have concluded that the splash transition can be identified by a critical Weber number, which we indeed find in the experiments. We have also found that the air cushion effect also sets the maximum deceleration experienced by the rim edge and have used this information to scale the sizes of the drops ejected for those experimental conditions in which $\Weber>\Weber_{crit}\simeq 140$ and the impact Weber number is below the threshold for which surface sealing occurs. Finally, we have found the interesting result that the thickness of the air layer entrapped beneath the disc sets the characteristic length scale of the drops ejected when the impact Weber number exceeds the critical one.\\

\begin{acknowledgements}
IRP and DvdM acknowledge financial support by NWO. JMG thanks financial support by the Spanish Ministry of Education under Project DPI2011-28356-C03-01 and the Junta de Andaluc\'ia under project P08-TEP-03997. The Spanish projects have been partly financed through European funds.
\end{acknowledgements}

\bibliographystyle{jfm}
\bibliography{splash}
\end{document}